\documentstyle[tighten,preprint,eqsecnum,prd,aps,epsf]{revtex}  

\def\pythiakt{${\rm k}_\perp$}
\def\NLOkt2{${\langle {\rm k_T^2} \rangle}$}
\begin{document}
\draft
\preprint{FERMILAB-Pub-97/030-E} 
\title{
Production of charm mesons at high transverse momentum \\
in 515~GeV/$c$ $\pi^-$-nucleon collisions}
%
%
\author{                                                                        
L.~Apanasevich,$^{4}$
J.~Bacigalupi,$^{1}$
W.~Baker,$^{3}$
M.~Begel,$^{9}$
S.~Blusk,$^{8}$
C.~Bromberg,$^{4}$
P.~Chang,$^{5}$
B.~Choudhary,$^{2}$
W.~H.~Chung,$^{8}$
L.~de~Barbaro,$^{9}$
W.~DeSoi,$^{9}$
W.~D\l ugosz,$^{5}$
J.~Dunlea,$^{9}$
E.~Engels,~Jr.,$^{8}$
G.~Fanourakis,$^{9}$
T.~Ferbel,$^{9}$
J.~Ftacnik,$^{9}$
D.~Garelick,$^{5}$
G.~Ginther,$^{9}$
M.~Glaubman,$^{5}$
P.~Gutierrez,$^{6}$
K.~Hartman,$^{7}$
J.~Huston,$^{4}$
C.~Johnstone,$^{3}$
V.~Kapoor,$^{2}$
J.~Kuehler,$^{6}$
C.~Lirakis,$^{5}$
F.~Lobkowicz,$^{9}$
P.~Lukens,$^{3}$
S.~Mani,$^{1}$
J.~Mansour,$^{9}$
A.~Maul,$^{4}$
R.~Miller,$^{4}$
B.~Y.~Oh,$^{7}$
G.~Osborne,$^{9}$
D.~Pellett,$^{1}$
E.~Prebys,$^{9}$
R.~Roser,$^{9}$
P.~Shepard,$^{8}$
R.~Shivpuri,$^{2}$
D.~Skow,$^{3}$
P.~Slattery,$^{9}$
L.~Sorrell,$^{4}$
D.~Striley,$^{5}$
W.~Toothacker,$^{7}$
N.~Varelas,$^{9}$
D.~Weerasundara,$^{8}$
J.~J.~Whitmore,$^{7}$
T.~Yasuda,$^{5}$
C.~Yosef,$^{4}$
M.~Zieli\'{n}ski,$^{9}$
V.~Zutshi$^{2}$
\\
{~}\\
\centerline{(Fermilab E706 Collaboration)}
{~}\\
}                                                                               
\address{                                                                       
\centerline{$^{1}$University of California-Davis, Davis, California 95616}
\centerline{$^{2}$University of Delhi, Delhi, India 110007}
\centerline{$^{3}$Fermi National Accelerator Laboratory, Batavia,              
                   Illinois 60510}                                              
\centerline{$^{4}$Michigan State University, East Lansing, Michigan 48824}     
\centerline{$^{5}$Northeastern University, Boston, Massachusetts  02115}
\centerline{$^{6}$University of Oklahoma, Norman, Oklahoma  73019}
\centerline{$^{7}$Pennsylvania State University, University Park, 
		   Pennsylvania 16802}
\centerline{$^{8}$University of Pittsburgh, Pittsburgh, Pennsylvania 15260}
\centerline{$^{9}$University of Rochester, Rochester, New York 14627}          
}                                                                               
\date{\today}
\maketitle
\begin{abstract}
  We present results on the production of high transverse 
momentum charm mesons in collisions of 515~GeV/$c$ negative pions with 
beryllium and copper targets. The experiment recorded a large sample
of events containing high transverse momentum ($p_T$) showers 
detected in an electromagnetic  
calorimeter. From these data, a sample of charm mesons has been 
reconstructed via their decay into the fully charged $K\pi\pi$ mode.
A measurement of the single inclusive transverse momentum
distribution of charged $D$ mesons from 1 to 8~GeV/$c$
is presented. 
An extrapolation of the measured differential cross section 
yields an integrated $D^{\pm}$ cross section of 
$11.4\pm 2.7(stat)\pm 3.3(syst) \ \mu$b per nucleon for $D^{\pm}$
mesons with $x_F>0$. The data are compared with 
expectations based upon next-to-leading order perturbative QCD, 
as well as with 
results from {\sc pythia}. We also compare our integrated $D^{\pm}$
cross section with measurements from other experiments.
\end{abstract}


\narrowtext
\section{INTRODUCTION}

  Over the past decade, measurements of charm production in hadronic 
interactions have provided valuable tests of the applicability of 
perturbative QCD (pQCD) to the production of heavy 
quarks\cite{mangano5,dameri,appel,tavernier}. 
Quark-antiquark annihilation and gluon fusion are the leading order (LO)
contributors to charm quark hadroproduction.
The next-to-leading order (NLO) contributions to the cross sections
have been evaluated\cite{nason3,nason1,beenakker1,beenakker2}, and
are comparable to the leading-order contributions.
While the results of the NLO calculations can accommodate the 
charm cross sections observed in $\pi^-$-nucleon 
interactions\cite{na11,na32a,e769-new,na32,na16,na27,e653},
the large size of the NLO contributions is an indication that still higher 
order contributions may be significant.  Furthermore, NLO calculations 
of the total charm quark cross section exhibit 
significant sensitivity to the choice of input parameters, including the
charm quark mass ($m_c$), the renormalization and factorization scales,
as well as the parton distribution functions (PDF's). 
For example, the calculated charm cross section changes by a factor 
of $\approx$3 when the renormalization scale is varied from
$2m_c$ to $\frac{1}{2}$$m_c$.
Varying the charm quark mass from 1.2~GeV/$c^2$ to
1.8~GeV/$c^2$ changes the calculated charm
cross section by as much as an order of magnitude\cite{mangano4,vogt}. 
While there is significant theoretical uncertainty in the total
charm quark production cross section, there is less uncertainty
in the shapes of the differential distributions.  
The shapes of the LO and NLO single inclusive charm quark
distributions versus $p_T$ are rather similar, and their shapes
exhibit smaller sensitivity to variations in $m_c$ or 
the renormalization scale\cite{mangano4,vogtscale}.

   To compare measurements of charm hadron production 
to the results of pQCD calculations, the consequences of
hadronization of the produced charm quarks must be taken into
account.  The fragmentation of charm quarks into charm hadrons is 
inherently a low momentum transfer process, and is therefore 
currently beyond the domain of pQCD.  Nevertheless, the effects 
of fragmentation may be described phenomenologically by convoluting the 
partonic cross sections with a suitable fragmentation function.
One hopes to describe the hadronization 
of charm quarks via a universal, process-independent fragmentation 
function, such as the Peterson {\it et al.}\cite{peterson} form, as measured 
in $e^+e^-$ collisions.  Convoluting the NLO prediction for charm 
quark production with a fragmentation function results in a softening of the 
predicted $p_T$ spectrum of the charm hadrons relative to the charm quarks.
The Peterson {\it et al.} fragmented NLO result is softer than 
the measured $p_T$ spectrum for charm hadrons, and it has been observed
that the unfragmented NLO result for charm quarks reproduces the shape of
observed charm hadron $p_T$ spectra reasonably well in the 
kinematic range $x_F>0$ and $p_T<4$~GeV/$c$\cite{mangano5}.
One might expect additional nonperturbative effects, such as the intrinsic 
transverse momentum of the incoming partons, to have 
an impact on the measured differential distributions. 
Frixione 
{\it et al.}\cite{mangano5} noted that the Peterson {\it et al.} fragmented 
NLO calculations of the charm $p_T$ spectra can be brought into agreement with 
data from experiments E769 and WA82, provided the partons (in each hadron) 
are supplemented with an average squared intrinsic transverse 
momentum, \NLOkt2=2.0~(GeV/$c$)$^2$.

  It is also of interest to compare the data with a  
Monte Carlo (MC) simulation that incorporates a model of
the fragmentation process (such as the LUND string model\cite{lund-string}
as implemented in {\sc pythia}). Again, 
nonperturbative effects may need to be taken into account 
to match the experimental results. 
{\sc pythia} simulates the so-called ``leading particle effect'', 
which results in an enhanced forward production for $D$ mesons
whose light quark is a spectator valence quark from the incoming
beam particle.  In addition to initial state radiation effects, 
{\sc pythia} also includes additional intrinsic transverse 
momentum characterized by the parameter \pythiakt\cite{ktnote}.
Reasonable agreement has been achieved between a
{\sc pythia} MC simulation and the E769 data using a \pythiakt\  value
of $\approx$1~GeV/$c$\cite{kwan}.

  Fermilab experiment E706\cite{alver} was designed to study 
high $p_T$ phenomena, principally associated with direct 
photons\cite{owens,ferbel} and high $p_T$ jets\cite{jetp}.
Jets arising from large momentum transfer collisions are expected 
to be rich in high $p_T$
charm particles. The high $p_T$ requirement of the
E706 trigger enhances the fraction of selected events
containing charm by nearly an order of magnitude 
compared to a minimum bias trigger. Since the
selected events result from high transverse momentum interactions,  
the data constitute a unique sample in which to study
charm particles (other recent fixed target charm experiments 
frequently employed lower threshold or minimum bias triggers, 
and yielded a rich sample of mostly lower $p_T$ events). 
From a theoretical standpoint,
one might expect the pQCD calculations to become more reliable in
the kinematic range accessible to E706. This paper
presents results of a study of high $p_T$ charm particles
produced in 515~GeV/$c$ $\pi^-$-nucleon collisions.
The measured differential cross sections are compared to 
NLO pQCD calculations and results from a {\sc pythia} MC simulation. 
We also compare the
integrated $D^{\pm}$ cross section with NLO pQCD calculations and
other recent measurements.

\section{APPARATUS}

\begin{figure}
\epsfxsize=6.5 truein
\epsffile[0 72 612 720]{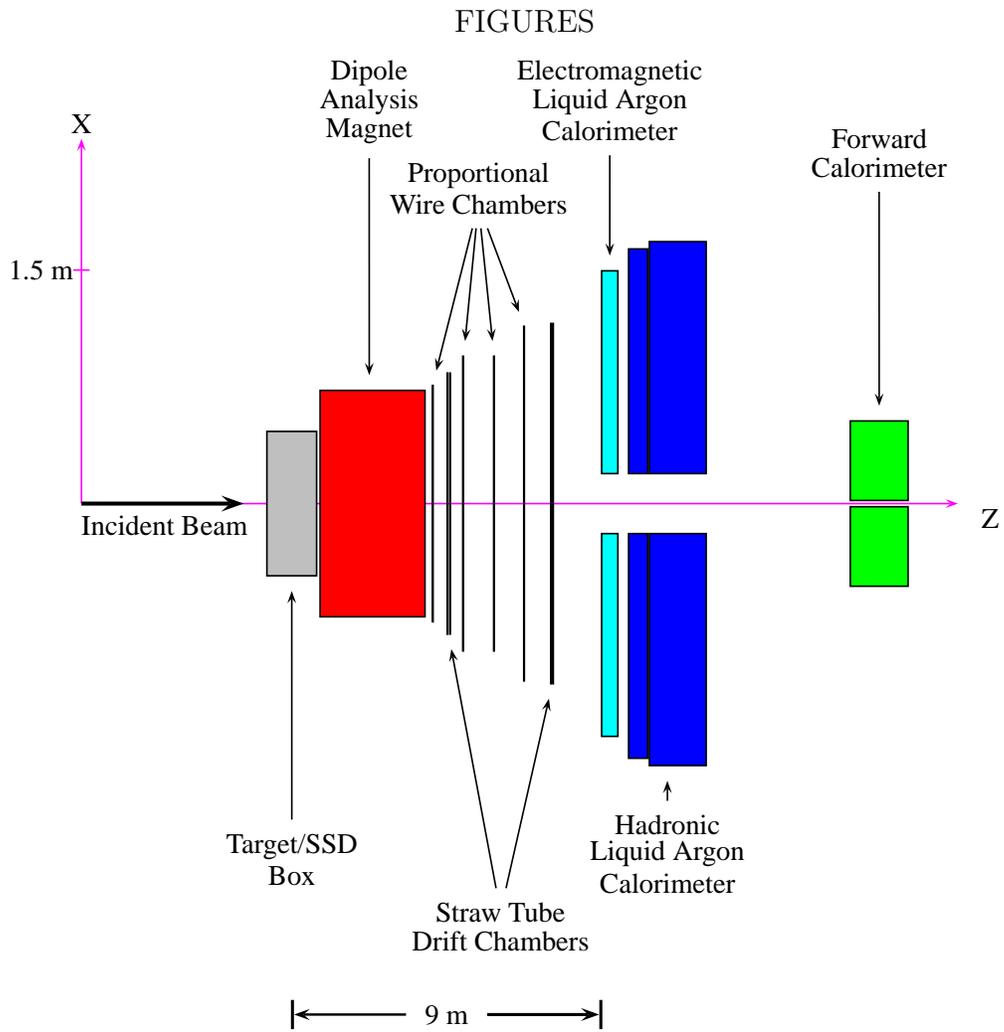}
\caption{Plan view of the 1990 configuration of the Meson West Spectrometer
(omitting muon identifiers).}
\label{spectrometer}
\end{figure}

  Experiment E706 was performed in the Fermilab Meson West beam line.
The unseparated negative secondary 515~GeV/$c$ beam was primarily
composed of pions with a small admixture of kaons ($<$5\%). 
Figure \ref{spectrometer} displays a
diagram of the key elements of the Meson West spectrometer
for this measurement\cite{e672-spec}. 
The detector included a precision charged particle 
tracking system and a large acceptance liquid argon 
calorimeter (LAC). The charged tracking system employed
silicon microstrip detectors (SSD), a large aperture dipole analysis 
magnet, proportional wire chambers (PWC), and
straw tube drift chambers (STDC). The LAC
contained a finely segmented electromagnetic section (EMLAC) as 
well as a hadronic section.  The high $p_T$ trigger was based upon
signals originating from showers detected in the EMLAC.
Only data from the charged particle 
tracking system and the electromagnetic section of 
the calorimeter contributed directly to this analysis.

\subsection{Charged particle tracking system}

\begin{figure}
\epsfxsize=6.5 truein
\epsffile[0 72 612 720]{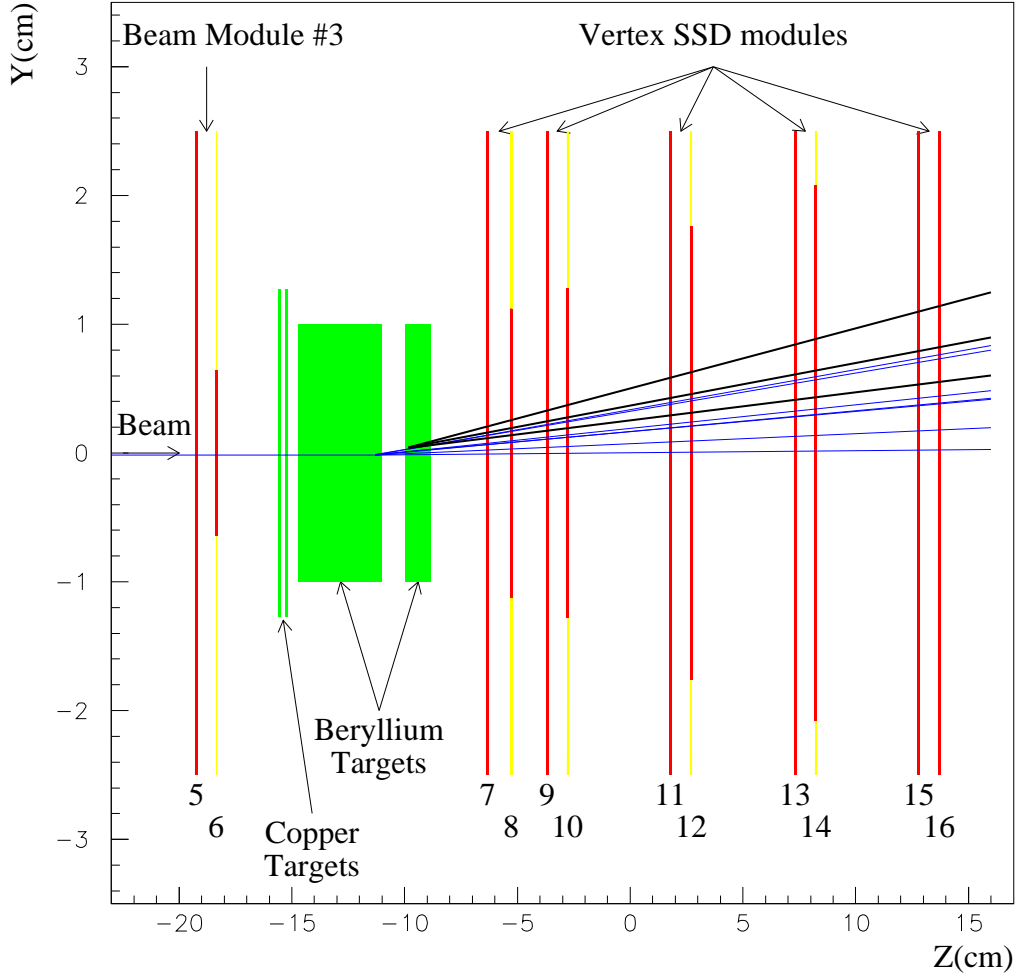}
\caption{Elevation view showing the configuration of the target and SSD 
region during the 1990 fixed target run.
Shown from left to right is the third of 3 beam SSD modules
(labelled as SSD planes 5 and 6), the copper and beryllium targets, and 
the 5 vertex SSD modules (labelled as SSD planes 7 through 16). The
instrumented regions of the SSDs are designated by the shaded regions.
The dotted lines illustrate the size of the SSD planes.
The odd numbered planes measure $X$ coordinates while the even numbered
planes measure $Y$ coordinates.  
The strips are 50~$\mu$m wide on all SSD planes except for the 
center $\pm$4.8~mm of SSD planes 7 and 8, where the strips are 
25~$\mu$m wide.
Reconstructed tracks from an interaction which includes a candidate 
charm particle are also depicted in the figure.}
\label{ssdpic}
\end{figure}

  The target region of the Meson West spectrometer (shown
in Fig.\ \ref{ssdpic}) consisted of nuclear targets and a 
SSD system\cite{engels-ssd}. The targets included two
780~$\mu$m thick copper pieces followed by two beryllium 
cylinders of length 3.71 cm and 1.12 cm, respectively. A
1 cm air gap between the two beryllium cylinders was 
designed for use in searches for heavy quark decays\cite{jesik}.
The targets were supported in a Rohacell stand
which had a cylindrical hole bored along the beam axis where 
the targets were positioned. 
Three beam SSD modules were located upstream of the target
and 5 vertex SSD modules were located downstream of the target.
Each module was composed of a pair of single-sided SSD planes
with strips aligned vertically and horizontally along the
$X$ and $Y$ axes, respectively.
All of the SSD planes were 50~$\mu$m pitch detectors, with the exception
of the  pair of vertex SSD's nearest the target, which consisted of
a high resolution 25~$\mu$m pitch central region and 50~$\mu$m
pitch outer regions. A total of 8912 
strips were instrumented, providing an angular acceptance of 
$\approx\pm$125~mrad in each view. 

   A large dipole analysis magnet was located downstream of the SSD system. 
Charged particles passing
through the magnet received a $p_T$ impulse of
$\approx$450~MeV/$c$ in the horizontal plane. Four PWC 
stations\cite{hartman,brown} were located 
downstream of this dipole analysis magnet, 
each separated by $\approx$1~m. Each station consisted 
of four proportional wire sense planes with wires oriented at the
angles $-90^{\circ}$ ($X$ view), $0^{\circ}$ ($Y$ view), $37^{\circ}$ 
($U$ view), and $-53^{\circ}$ ($V$ view). Thus, the $X$ and $Y$ views 
were orthogonal to one 
another, as were the $U$ and $V$ views. Within each sense wire plane, the 
spacing between the sense wires was 0.254~cm.  A total of 13,440 sense wires 
were instrumented. The STDC system\cite{bromberg} consisted of two
stations, with each station consisting of 4 $X$ view planes 
followed by 4 $Y$ view planes. The $X$ and $Y$ view planes in the upstream 
station consisted of 160 and 128 tubes, respectively.  Each 
tube had a diameter of 1.03~cm.  Each plane of the downstream STDC 
consisted of 168~tubes, and each tube had a diameter of
1.59~cm. The hit resolution of individual tubes in the STDC's 
was typically in the range of 200 to 300~$\mu$m.

\subsection{Electromagnetic calorimeter}

  The EMLAC\cite{lobk,desoi}, which was located 9~m downstream of 
the target, was assembled from four independent quadrants, each of which
instrumented $\frac{1}{4}$ of the azimuthal acceptance of the detector.  
The inner and outer radii
of the EMLAC were 20 cm and 160 cm respectively, corresponding to a center 
of mass rapidity coverage of $-1$ to 1 (for the incident 515~GeV/$c$
beam).  Each quadrant was composed of 
66 layers.  Each layer consisted of an absorber sheet (which was 2~mm of lead 
in all but the first layer), a 2.5~mm liquid argon gap, a pair of octant-size 
copper-clad G-10 anode boards, and another 2.5~mm argon gap.  In alternating
layers, the copper cladding of the G-10 anode boards was cut to form 
either concentric (R) strips or azimuthal ($\phi$) strips.  The locations and 
widths of the R strips on the radial anode boards were such that the R strips 
were focussed on the target region so that neutral particles produced in the 
target passed through the same R strip in each successive radial 
anode board.  The width of the R strips on the first anode board was 5.5~mm. 
The interleaved azimuthal anode boards were subdivided at a radius of 40~cm 
into inner $\phi$ and outer $\phi$ regions.  Each of the inner $\phi$ strips 
subtended an angle of $\frac{\pi}{192}$ radians in azimuth while the outer 
$\phi$ strips subtended an angle of $\frac{\pi}{384}$ radians.  
Longitudinally, the EMLAC was read out in two sections.  The front section 
consisted of the first 22 layers (8.5 radiation lengths) while 
the back consisted of the remaining 44 layers (18 radiation lengths).
For each section in each octant, signals from corresponding R ($\phi$) 
strips were ganged together and read out independently.

\subsection{Trigger}

   The trigger selected events producing high transverse 
momentum showers in the EMLAC. This event selection process involved four 
stages; beam and interaction definitions,
a pretrigger requirement, and the final trigger requirements\cite{sorrell}.
The beam definition required that a single beam particle
was detected in the beam hodoscope located 2 m
upstream of the target. 
A scintillation counter with a $\frac{3}{8}$ inch diameter
hole was located downstream of
the beam hodoscope and was used to reject interactions initiated 
by particles in the beam halo.
Pairs of interaction counters were mounted near the upstream and 
downstream mirror plates of the dipole analysis magnet.
An interaction was defined as a
coincidence between signals from at least two of these four 
interaction counters. To minimize potential confusion due to out-of-time 
interactions, a cleaning filter 
rejected any interactions that occurred within $\pm$60~ns
of each other.  For those interactions that satisfied 
the beam and interaction definition, the $p_T$ deposited in
various regions of the EMLAC was evaluated by weighting 
the energy signals from the EMLAC R channel amplifier fast outputs
by $\approx\sin{\theta_i}$, where $\theta_i$ is the polar angle 
that the $i^{th}$ strip subtends with respect to the nominal beam axis. 
The pretrigger $p_T$ requirement was satisfied if the
$p_T$ detected in the inner 128 R channels or the outer 128 R
channels of at least one octant was greater than a threshold
of $\approx$1.7~GeV/$c$ (for the PRETRIGGER HI). 
A pretrigger signal was issued only if the signals
from a given octant satisfied that pretrigger $p_T$ requirement
and there was no evidence in that octant of substantial noise or 
significant $p_T$ attributable to an earlier interaction and there
was no incident beam halo muon detected\cite{muon-prot}. The pretrigger
signal latched the data from the various subsystems while the
final trigger decision was being evaluated.

   The experiment employed several different high $p_T$ trigger 
definitions that were based upon the LOCAL and 
GLOBAL signals from octants that satisfied the pretrigger. 
Local trigger groups were formed by clustering the 256 R channels 
in each octant into 32 groups of 8 channels. Each of the
adjacent pairs of groups of 8 channels (groups 1\&2, 2\&3, $\ldots$~,\  31\&32)
formed a local group of 16 strips. If the $p_T$ detected in 
any of these groups of 16 was above a specified high (or low) threshold,
then a LOCAL HI (or LO) signal was generated for 
that octant. 
A GLOBAL HI (or LO)
signal for a given octant was generated if the summed $p_T$
from the groups of 8 in that octant was above a specified high (or low)
threshold. In order to suppress coherent noise effects, only groups
of 8 registering at least $\approx$250~MeV/$c$ contributed to this global
$p_T$ sum. (This cutoff was applied independently to signals from 
groups of 8 from the front and back sections of the EMLAC.)
Three of the trigger types, which accounted for $\approx$80\% of the 
E706 data, were used in this charm analysis; they
were the SINGLE LOCAL HI, the LOCAL$\otimes$GLOBAL HI, and the 
TWO-GAMMA triggers. The SINGLE LOCAL HI trigger required
a LOCAL HI signal from an octant that satisfied the PRETRIGGER HI.
The LOCAL HI threshold was $\approx$3~GeV/$c$. The LOCAL$\otimes$GLOBAL HI
trigger required the coincidence of a LOCAL LO signal (threshold 
$\approx$1.7~GeV/$c$),
a GLOBAL HI signal (threshold $\approx$3~GeV/$c$), and a PRETRIGGER HI all 
from the same octant\cite{dptrig}. The TWO-GAMMA trigger required LOCAL LO
signals from any two octants that were separated by at least
90$^{\circ}$ in azimuth, where both octants also satisfied the lower
threshold PRETRIGGER LO requirement.

    In addition to the high $p_T$ triggers, a prescaled sample of
low bias triggers were recorded concurrently.
These low bias triggers included beam,  
interaction and pretrigger events,
and constituted $\approx$10\% of the recorded 
events.

\section{CHARGED TRACK RECONSTRUCTION}

   The charged tracks in the selected events were reconstructed 
first in the PWC system.
Since the PWC system consisted of four views, three-dimensional (3D)
information was extracted. Space tracks were formed 
by combining all possible $XY$ ($UV$) candidate track segments and
searching for hits along the projections in the remaining 
two views. Space tracks were required to have a minimum of 11 hits
if they involved all four PWC stations (16 planes), 10 hits if they 
involved three PWC stations (12 planes), and 6 hits if they involved 
only the two upstream PWC stations (8 planes). 
 
  The STDC pattern recognition was initially seeded by
the 3D space tracks reconstructed in the PWC system (in order to 
correlate the hits in the $X$ and $Y$ views of the straw tubes).
Once the correlation was performed, an iterative procedure was
used to form straw tracks using only the hits detected in the STDC's.
Straw track segments
required a minimum of 4 hits in either the $X$ or $Y$ view.
The angular resolution of straw tracks was $\approx$0.06~mrad.
After reconstructing the straw track segments, each space track was refit
using the hit information from both the PWC and STDC systems.

   After identifying the downstream space tracks, track
segments were reconstructed in the $X$ and $Y$ views of the SSD
system. Four and five hit tracks
were reconstructed, and then three hit tracks were formed from the
previously unused hits. The SSD tracks had an average
angular resolution that was similar to the STDC's, and an 
impact parameter resolution at the primary vertex of $\approx$15~$\mu$m (for 
$p \gtrsim 15$~GeV/$c$). Track segments were also reconstructed from
the hits in the beam SSD modules to measure the direction 
of the incident beam particle.

\begin{figure}
\epsfxsize=6.5 truein
\epsffile[0 72 612 720]{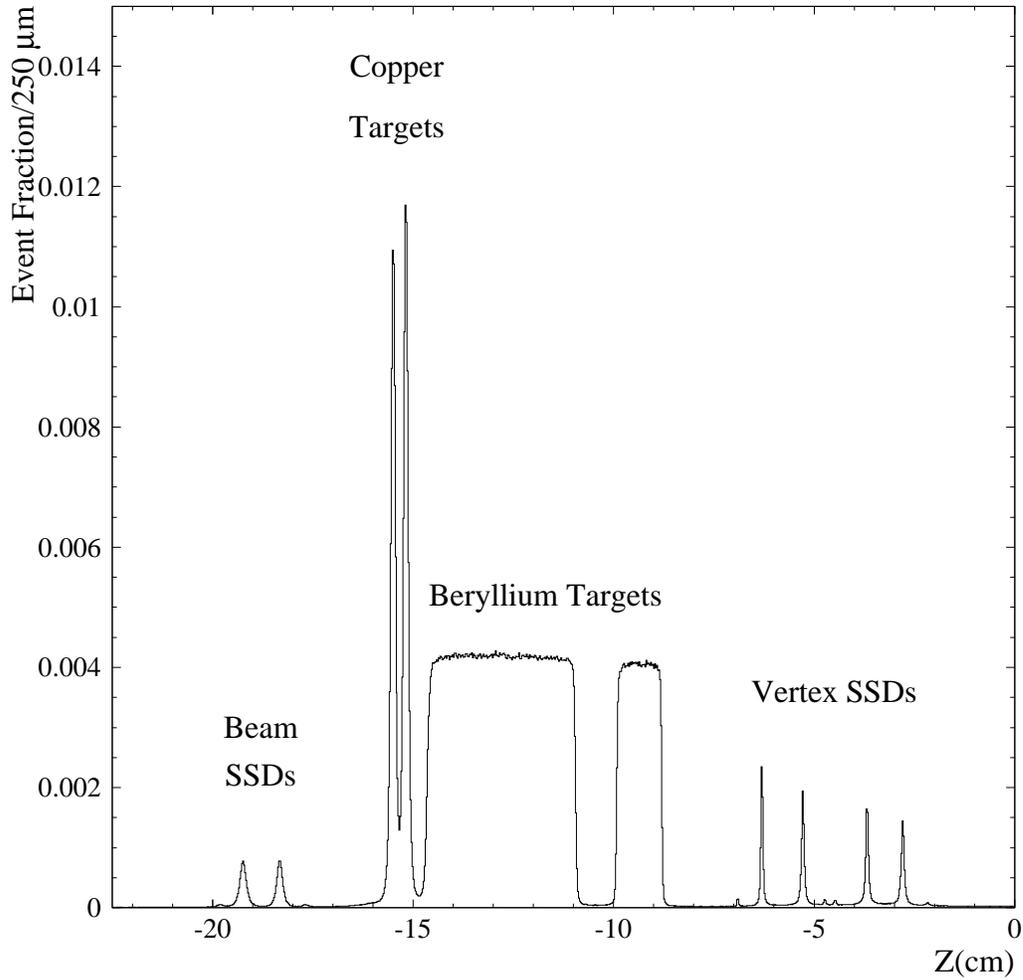}
\caption{Longitudinal positions of the reconstructed primary vertices for a
sample of events containing high $p_T$ showers
acquired during the 1990 fixed target run. This distribution is 
not corrected for beam attenuation.}
\label{vertex}
\end{figure}

   Three dimensional tracks were formed in the SSD system
by linking the projected 3D downstream (PWC and STDC) tracks 
to corresponding projected
SSD $X$ and $Y$ track segments at the center of the analysis magnet. 
In the bend plane of the analysis magnet, a link between an SSD $X$
track segment and a downstream space track was established
if the corrected difference between the $X$ positions at the magnet center
of the projected SSD X track segment and the projected downstream track
was within 3.3$\sigma_{\Delta X}$. A link between an SSD $Y$ track segment
and a downstream space track occurred when the corrected difference between
the $Y$ positions of the projected tracks at the magnet center was
within 3.3$\sigma_{\Delta Y}$ and the corrected slope difference
between the tracks was within 3.3$\sigma_{\Delta\theta_Y}$ \cite{corr}.
The matching resolutions for
the projections ($\sigma_{\Delta X}$ and $\sigma_{\Delta Y}$) and
the $YZ$ slopes ($\sigma_{\Delta\theta_Y}$) were momentum dependent
functions which were extracted from the data. SSD $X$ and $Y$ view track
segments were correlated with each other by virtue of being linked
to the same downstream space track. Once the linking was 
complete, the primary vertex for the event was reconstructed\cite{blusk}.
Figure \ref{vertex} shows a distribution of reconstructed
primary vertex locations along the nominal beam direction ($Z$ axis) for 
events accumulated during the 1990 fixed target run. Based upon
the primary vertex location, one can identify whether the
interaction occurred in the beryllium, copper, or silicon (SSD)
targets. The average resolution for the $Z$ location of the primary vertex 
was $\approx$300~$\mu$m. After the primary vertex was located,
the direction cosines, charge, and momentum of each reconstructed 
charged track were evaluated. (The momentum scale was calibrated using
$J/\psi$ and $K^0_S$ signals.) 
The average momentum resolution for charged tracks produced in
the target region was $\sigma_p/p \approx 0.0076 + 0.00026p$, 
where $p$ is the momentum measured in GeV/$c$. 

   After the full sample of events was reconstructed and the results 
written in the form of data summary tapes (DSTs), a secondary vertex
analysis was performed to search for evidence of heavy quark
production. Only events that had a reconstructed primary vertex
in the copper or beryllium targets were used in this analysis
($-17.1$~cm~$<~Z~<$~$-8.5$~cm). For each event, tracks having large
{\it transverse significance\/} to the primary vertex 
were identified as secondary tracks. Transverse significance 
is defined as the measured impact parameter between a vertex and
a given track candidate divided by the corresponding expected
uncertainty. The algorithm evaluated
all pairs of secondary tracks and selected only those
pairs that were consistent with emanating from the same space 
point. For all such combinations, the algorithm 
determined whether any other secondary tracks had transverse
significance of less than 3 units relative to the space point in question.
All such tracks were added to the
track list associated with that secondary vertex, and the vertex 
location and its associated uncertainties were reevaluated 
(via a $\chi^2$ minimization technique).
A list of vertices, each determined by two or more space tracks was 
thus generated. Two track vertices were
referred to as {\it vees\/} and all other vertices were referred to as
{\it secondary vertices}. To minimize the losses
introduced by only utilizing secondary tracks,
all space tracks in the event were examined to determine
whether they might belong to any given secondary vertex or vee.
This phase of the program generated a list of additional tracks 
that might possibly belong to each of the secondary vertices or vees.
Neither the secondary vertices nor the vees were refit with any of
these additional tracks. Only those events with at least one 
reconstructed secondary vertex or vee contributed to this analysis.

\section{CHARM SIGNALS IN THE E706 DATA}

\begin{figure}
\epsfxsize=6.5 truein
\epsffile[0 72 612 720]{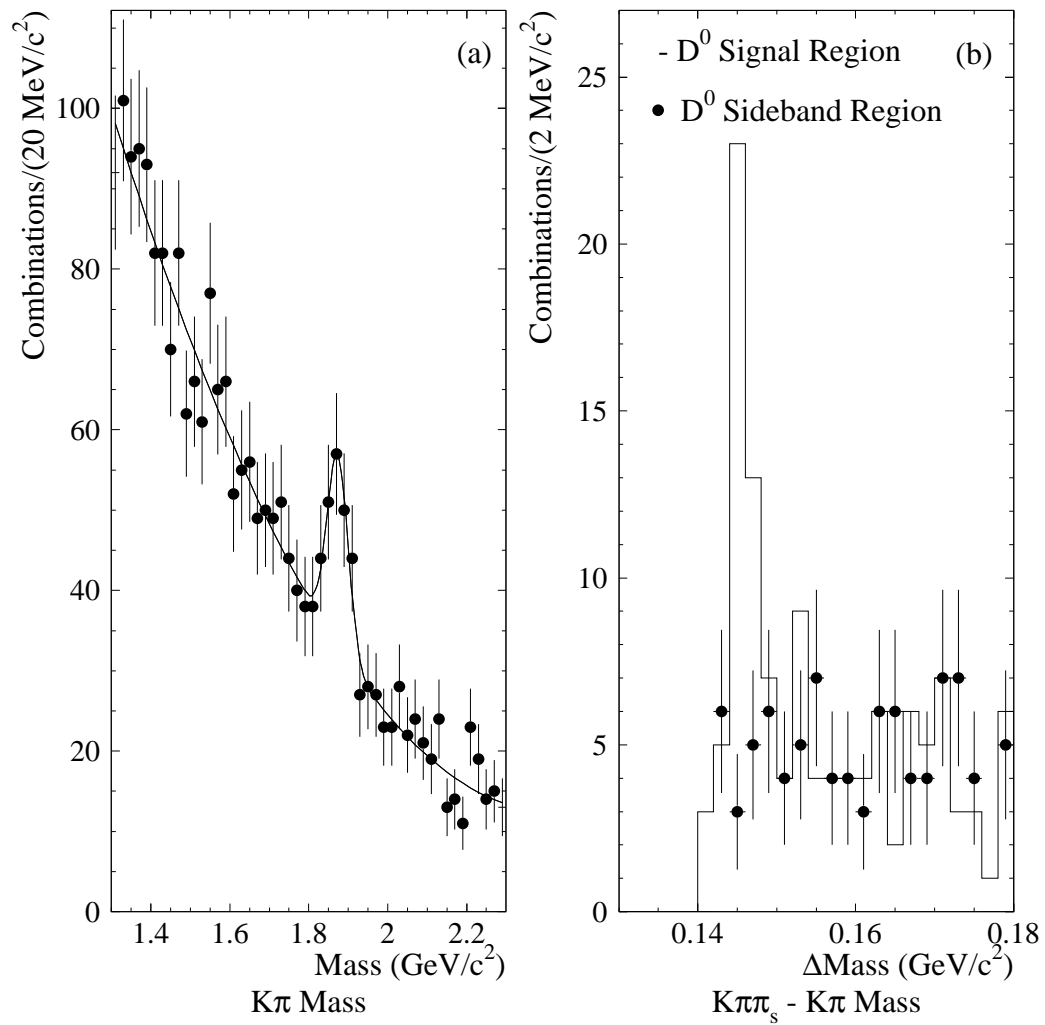}
\caption{(a) The combined $K^-\pi^+$ and $K^+\pi^-$ invariant mass distribution
for vees (two track secondary vertices) with $p_T>1$~GeV/$c$, and (b) the mass 
difference between
$K\pi\pi_s$ and $K\pi$ combinations for the signal and sideband regions of the 
neutral $D\to K\pi$ candidates. The $\pi_s$
is a relatively low momentum (soft) pion that is attached to the  primary 
interaction vertex.}
\label{d0-dstar}
\end{figure}

   From the data sample acquired during the 1990 fixed target run,
we have identified $D^0$, $D^{*\pm}$,  
and $D^{\pm}$ signals in fully charged modes. The $D^0$ and $D^{*\pm}$
signals (see Fig.\ \ref{d0-dstar}) do not directly contribute to the
measurements presented in this report, which are
based upon the sample of $D^{\pm}$ mesons that were observed
via their decay to the fully charged final
state $K^{\mp}\pi^{\pm}\pi^{\pm}$. The $D^{\pm}$ sample was extracted
from the subset of events that contained at least one
secondary vertex with 3 tracks or a vee with additional tracks
attached to it. For all such vertices (vees), the three-body
invariant mass was evaluated by assigning the charged pion mass to each of 
the two like-charge tracks
while the oppositely charged track was assigned the kaon mass.
To reduce the large combinatorial background, only secondary 
vertices that satisfied additional requirements contributed to 
the final analysis. The significant requirements were:
(1) the impact parameter to the primary vertex of the parent momentum vector
formed from the candidate decay products 
must be less than 50~$\mu$m; (2) the longitudinal separation
between the primary and secondary vertex normalized by the
corresponding expected uncertainty in that separation must be 
at least six; (3) the impact parameter of each candidate decay track
to the secondary vertex (vee) must be less than 0.4 of the corresponding 
impact parameter to the primary vertex for every candidate decay track
that contributed to the determination of the secondary vertex (vee)
location; (4) for those tracks that contributed to the secondary vertex
finding, the product of the secondary to primary vertex impact
parameter ratios must be less than 0.005 for secondary vertices or
less than 0.002 for vees.

\begin{figure}
\epsfxsize=6.5 truein
\epsffile[0 72 612 720]{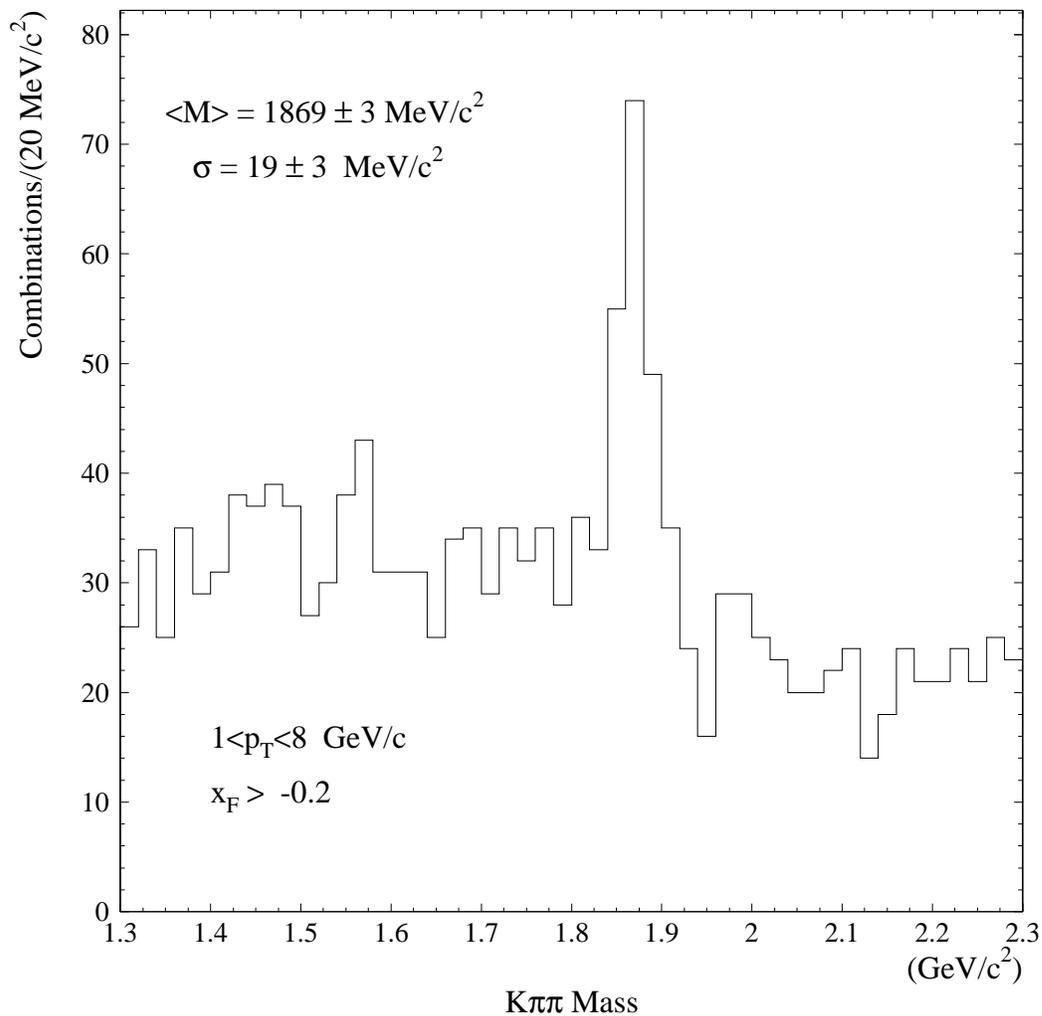}
\caption{The $K^{\mp}\pi^{\pm}\pi^{\pm}$ invariant mass spectrum for those
events satisfying
all reconstruction requirements from the 1990 fixed target run 
data sample. All contributing $K\pi\pi$ combinations have
$x_F>-0.2$ and $p_T>1$~GeV/$c$.}
\label{signal-allxf}
\end{figure}

\begin{figure}
\epsfxsize=6.5 truein
\epsffile[0 72 612 720]{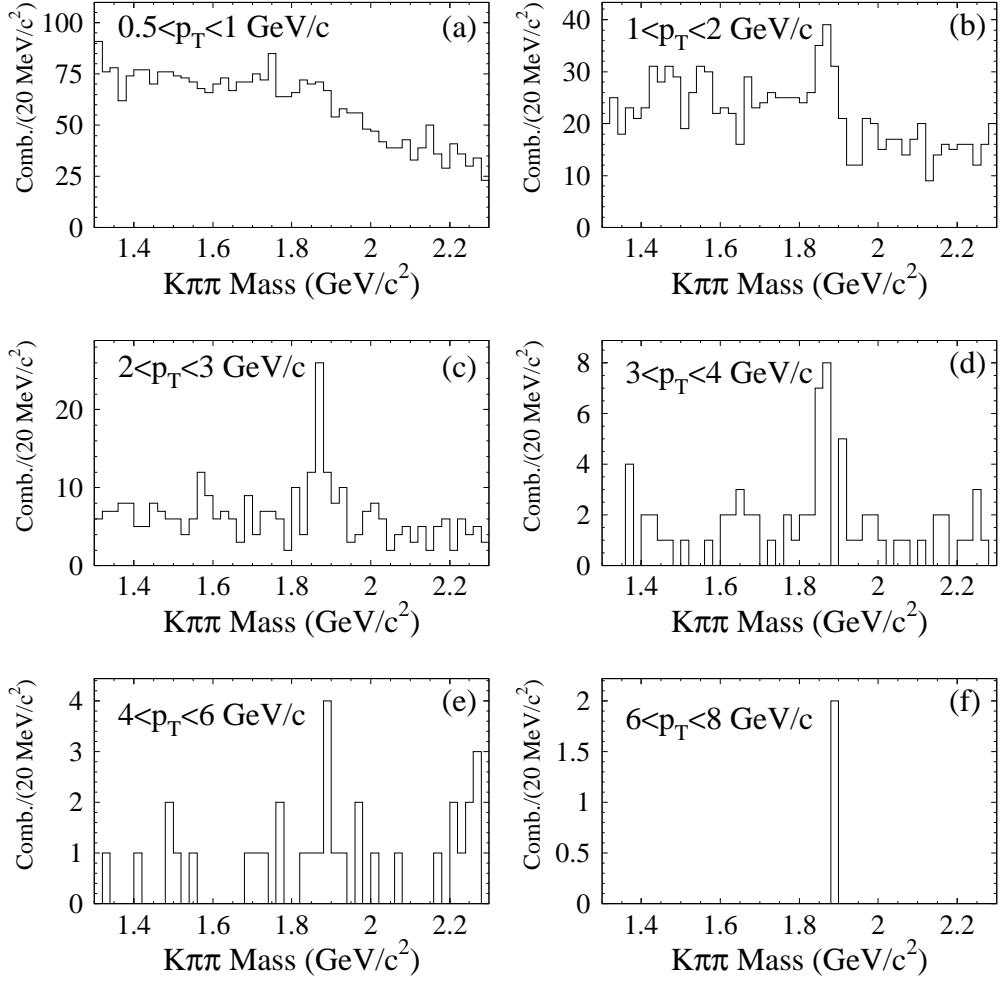}
\caption{The $K^{\mp}\pi^{\pm}\pi^{\pm}$ invariant mass spectrum in 
$p_T$ intervals. The $p_T$ intervals are (a) 0.5 to 1~GeV/$c$, 
(b) 1 to 2~GeV/$c$, (c) 2 to 3~GeV/$c$, (d) 3 to 4~GeV/$c$, (e) 4 to 6~GeV/$c$, 
and (f) 6 to 8~GeV/$c$ as indicated. The distributions are integrated over 
the region $x_F>-0.2$.}
\label{signal-pt-allxf}
\end{figure}

  Figure \ref{signal-allxf} shows the $K\pi\pi$ invariant mass
spectrum from the events satisfying the above analysis requirements.
The mass distribution
contains 110 $\pm$ 17 combinations above background in
the $D^{\pm}$ mass region (1.8 to 1.94~GeV/$c^2$) with
$x_F\equiv 2p_Z/\sqrt{s} >-0.2$ and $p_T>1$~GeV/$c$.
Figure \ref{signal-pt-allxf}, which shows mass
spectra in several $p_T$ intervals, illustrates the broad $p_T$
range populated by this $D$ sample. This large range
is a consequence of the E706 trigger, which preferentially selected 
high $p_T$ interactions. The high trigger
thresholds coupled with the large combinatorial background
compromise our ability to observe a significant signal
for $D$ mesons with $p_T<1$~GeV/$c$ 
[as indicated in Fig.\ \ref{signal-pt-allxf}(a)].
(The main source of background is secondary 
interactions in the target.) The $D$ signal region was defined 
to be between 1.80 and 1.94~GeV/$c^2$, except for the interval
$1<p_T<2$~GeV/$c$, where a narrower mass range of 1.82 to 1.92~GeV/$c^2$ 
was employed. This more restrictive mass requirement reduces the
statistical uncertainty with only a minimal loss of efficiency.
Since the resolution of the reconstructed $D$ signal ($\approx$19~MeV) 
observed in the data and our MC simulation were consistent, the small 
loss of events resulting from the narrower mass range was absorbed into
the reconstruction efficiency. The background in the signal
region was estimated via a linear interpolation between the lower and
upper sideband regions. The uncertainty in the number of background 
combinations was estimated by fitting the background to first and second 
order polynomials over several mass regions (which included the signal region).

\begin{figure}
\epsfxsize=6.5 truein
\epsffile[0 72 612 720]{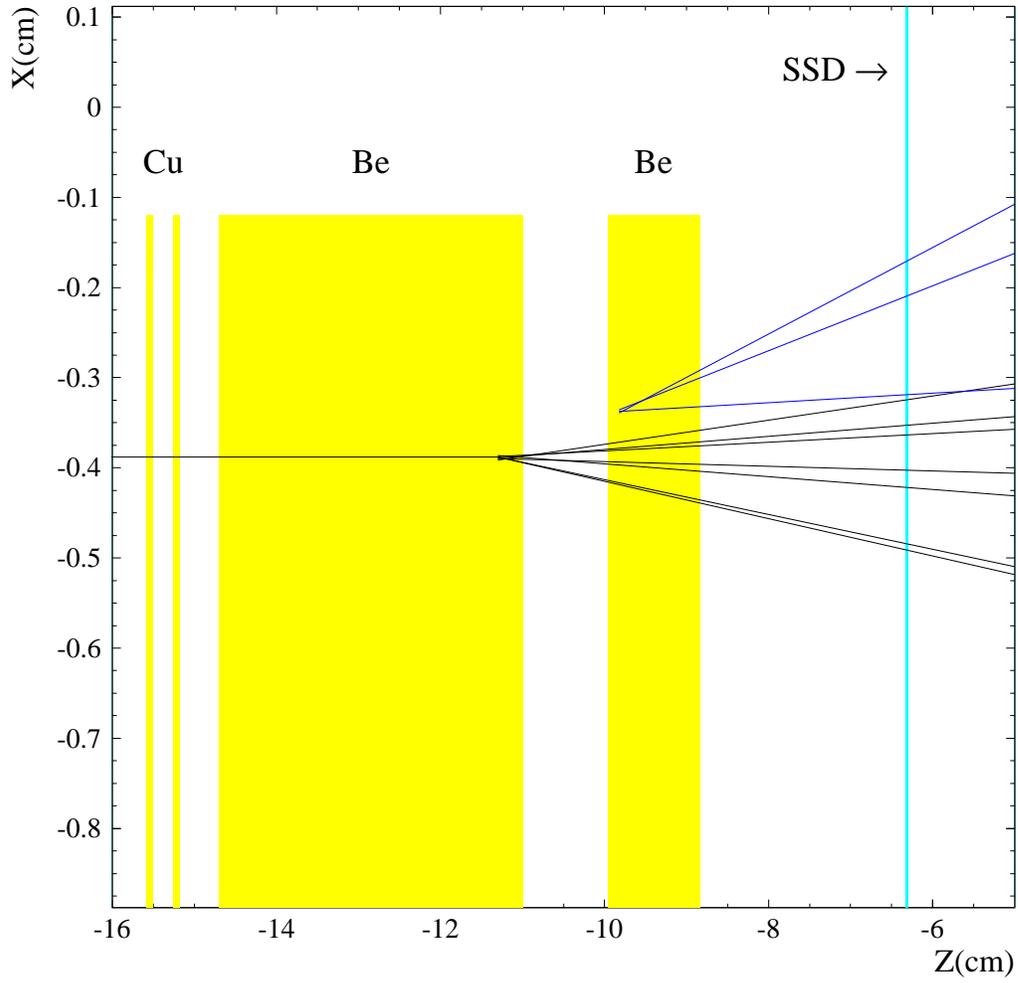}
\caption{A illustration of the reconstructed tracks in the
$XZ$ view of an event containing a candidate high $p_T$ charged $D$
meson decaying into the
fully charged $K\pi\pi$ mode downstream of the primary vertex.}
\label{event_x}
\end{figure}

   Figure \ref{event_x} shows the tracks from an event containing a 
reconstructed charm particle candidate with a $p_T$ of
4.1~GeV/$c$. The figure shows the various target elements, the
first vertex SSD plane, and the projected charged 
tracks from the event which reveal a primary vertex and a displaced
three-track secondary vertex. Due to the high transverse
momentum of the charm candidate, the secondary vertex is 
well isolated from the other charged tracks in the event. 
Note that the vertical scale is magnified
with respect to the horizontal scale; the polar angle of the 
widest angle track is no more than about $6^{\circ}$.

\section{EFFICIENCIES}

   An event simulation was used to estimate the efficiency
for selecting events containing charged $D$ mesons and for
detecting those $D^{\pm}$ decays. The simulation
includes an event generator and a detector simulation.
The event generator simulates particle production in
high energy collisions, and the detector simulation
models the response of the detectors to the generated particles.
The details of this simulation and the evaluation 
of the efficiencies are discussed below.

\subsection{Event generation}

   The event generator chosen to produce
full events was the {\sc pythia} 5.6/{\sc jetset} 7.3 
package\cite{pylund}.  The physics processes employed
in the event generation are specified by the user. The physics
processes investigated in this analysis included minimum bias 
events and a pure charm event sample. The former was used to tune the
MC parameters to match the global characteristics of events observed 
in the data. Once the MC simulation was tuned, the efficiencies
for triggering on and selecting events containing charm particles
were evaluated using the pure charm event sample.
{\sc pythia} describes
the hard scattering between hadrons via leading order 
perturbative QCD matrix elements, and simulates the NLO 
contributions through effective K factors\cite{pylund,kfactor}.
Parton showers are produced via perturbative branchings
of one parton into two or more partons. The {\sc pythia} simulation 
also includes initial state radiation of the incoming partons,
which by default, is activated. {\sc jetset} handles the
nonperturbative fragmentation of the final state colored
partons into colorless hadrons using the LUND string 
model\cite{lund-string}. In addition, {\sc jetset} handles the 
decays of unstable particles via a list of decay modes and 
branching ratios that are extracted from  
Particle Data Group tables. For each event,
the event generator provides a list of the resulting stable 
particles (and their associated kinematic variables) that
can be used as input to the E706 detector simulation. The
detector simulation is based upon the {\sc geant} 
software package\cite{geant}.

\begin{figure}
\epsfxsize=6.5 truein
\epsffile[0 72 612 720]{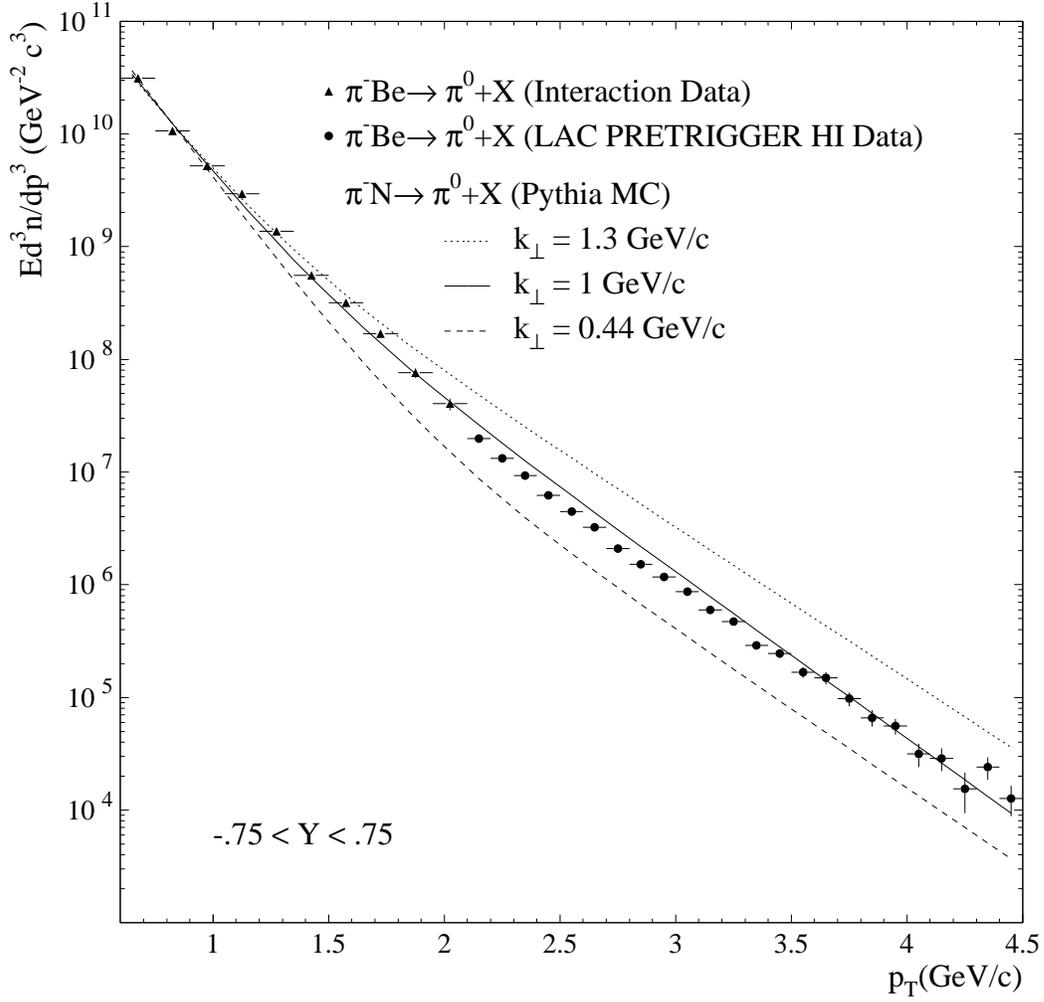}
\caption{Corrected $\pi^0$ production spectra from the 
E706 data and the {\sc pythia} MC event generator. The MC spectra are
shown for several choices of the {\sc pythia} intrinsic parton transverse 
momentum parameter \pythiakt.} 
\label{pi0pt}
\end{figure}

\begin{figure}
\epsfxsize=6.5 truein
\epsffile[0 72 612 720]{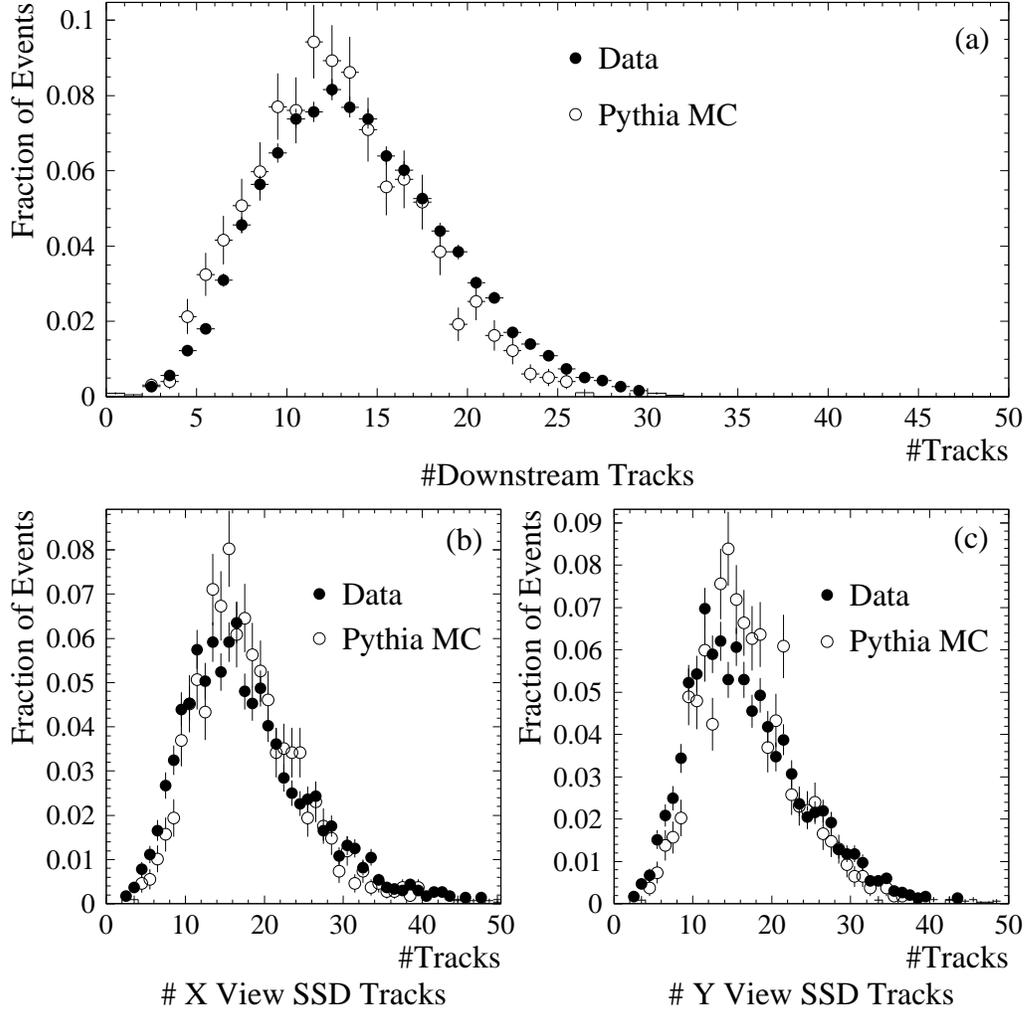}
\caption{Normalized distribution of the number of reconstructed tracks 
in (a) the downstream tracking system,
(b) the SSD $X$ view, and (c) the SSD $Y$ view for events satisfying at
least one of the triggers used in this analysis. The filled circles 
represent the data and the open circles are the corresponding results 
from the MC simulation.}
\label{trks}
\end{figure}

   The event generator was first tuned to match various
distributions observed in our data which were relevant to this 
analysis. Since the trigger discriminated using electromagnetic
depositions, it is important to reproduce the
$p_T$ spectrum of particles that produce electromagnetic showers. 
Figure \ref{pi0pt} shows
the spectrum of $\pi^0$'s measured in the data and the
corresponding spectra generated by {\sc pythia}
for several choices of the \pythiakt\  parameter. The data distribution 
was measured using the low bias triggers, and the MC spectrum was 
generated using minimum bias events.
In each case, the MC spectrum is normalized to the
same integral as the data over the kinematic range shown in the
figure. The data are reasonably described by the {\sc pythia} result
generated using a \pythiakt\  value of $\approx$1~GeV/$c$,
which is larger than the {\sc pythia} default \pythiakt\  value 
of 0.44~GeV/$c$.
The $p_T$ spectra of charged tracks were also compared and exhibited a
similar level of agreement. We chose a \pythiakt\  value of 1.0~GeV/$c$ for 
further study.
To improve the match between the charged particle
multiplicity observed  in our data and {\sc pythia}, we adjusted the
{\sc pythia} parameter designated as the {\it effective minimum 
transverse momentum for multiple interactions}\cite{par-tune},
which increased the mean track multiplicity by $\approx$30\%.
Figure \ref{trks} shows
the multiplicity distributions of charged tracks as reconstructed 
in the PWC system and in the $X$ and $Y$ views of the SSD system
for both the data sample and the Monte Carlo sample 
(after adjustment of the aforementioned {\sc pythia} parameter)
for events that satisfied at least one of the triggers used in this analysis.
The corresponding data and MC distributions are similar,
indicating that the adjusted MC simulates the particle multiplicity of
high energy collisions reasonably well. Rapidity distributions
of charged tracks in the MC and data were also found to be
in reasonable agreement. With
these modifications, the {\sc pythia} simulation adequately describes 
the distributions of final state particles observed in our data.

\subsection{Trigger simulation}

 As previously described, the E706 trigger utilized signals from
the electromagnetic section of the LAC. Since the EMLAC consists of $\approx$27 
radiation lengths, but only $\approx$1 interaction length, photons deposit 
nearly all of their energy in the EMLAC, whereas hadrons usually do not.
Consequently, high $p_T$ photons and electrons were more likely to
trigger the apparatus than hadrons of the same $p_T$.
To investigate the impact of the requirements imposed by
the trigger on observed events, we developed a software simulation
of the online trigger.  Corrections for the losses
resulting from the high $p_T$ thresholds of the various triggers
were evaluated by subjecting {\sc pythia} events to this
software trigger simulation.
The main features of the trigger simulation are discussed in
this section.

\begin{figure}
\epsfxsize=6.5 truein
\epsffile[0 72 612 720]{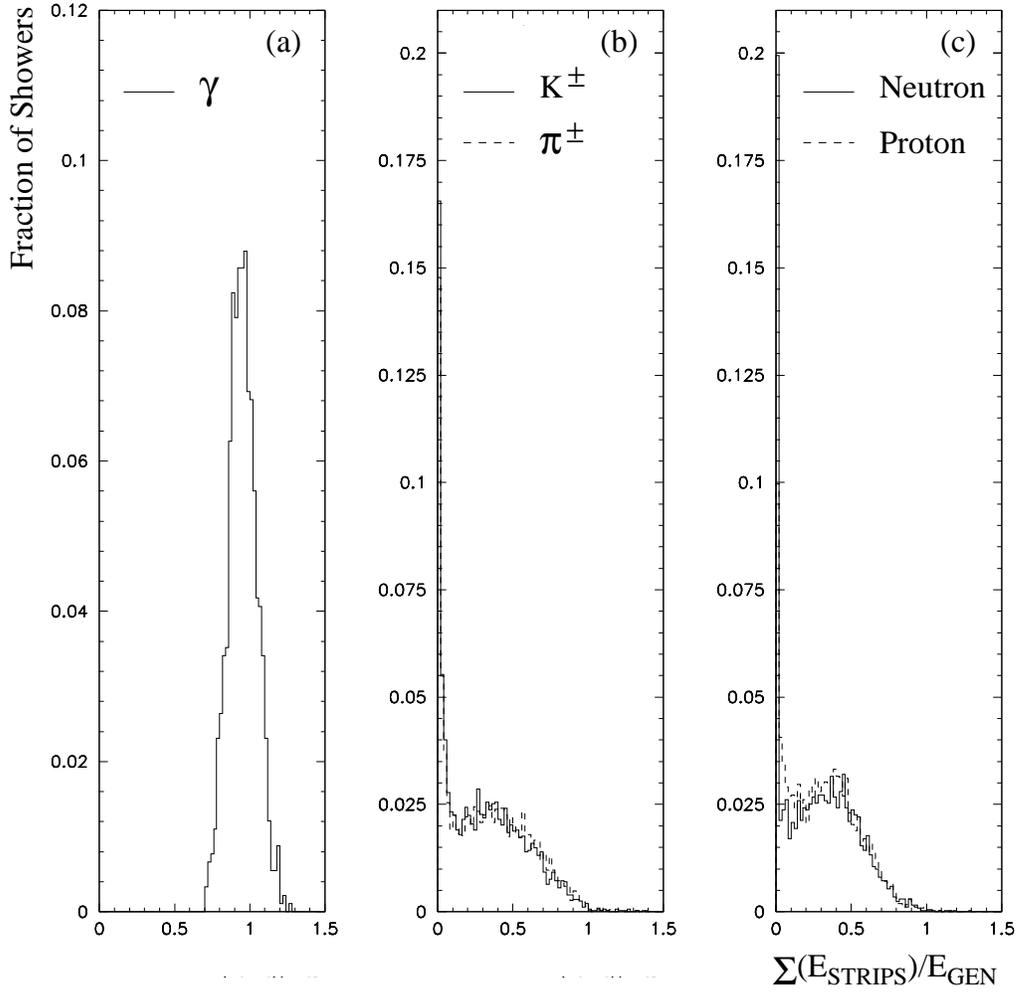}
\caption{Probability distributions of the calibrated energy
response of the EMLAC for simulated incident 20~GeV (a) photons, (b) mesons, 
and (c) baryons relative to the energy of the incident simulated particle.}
\label{eop}
\end{figure}

\begin{figure}
\epsfxsize=6.5 truein
\epsffile[0 72 612 720]{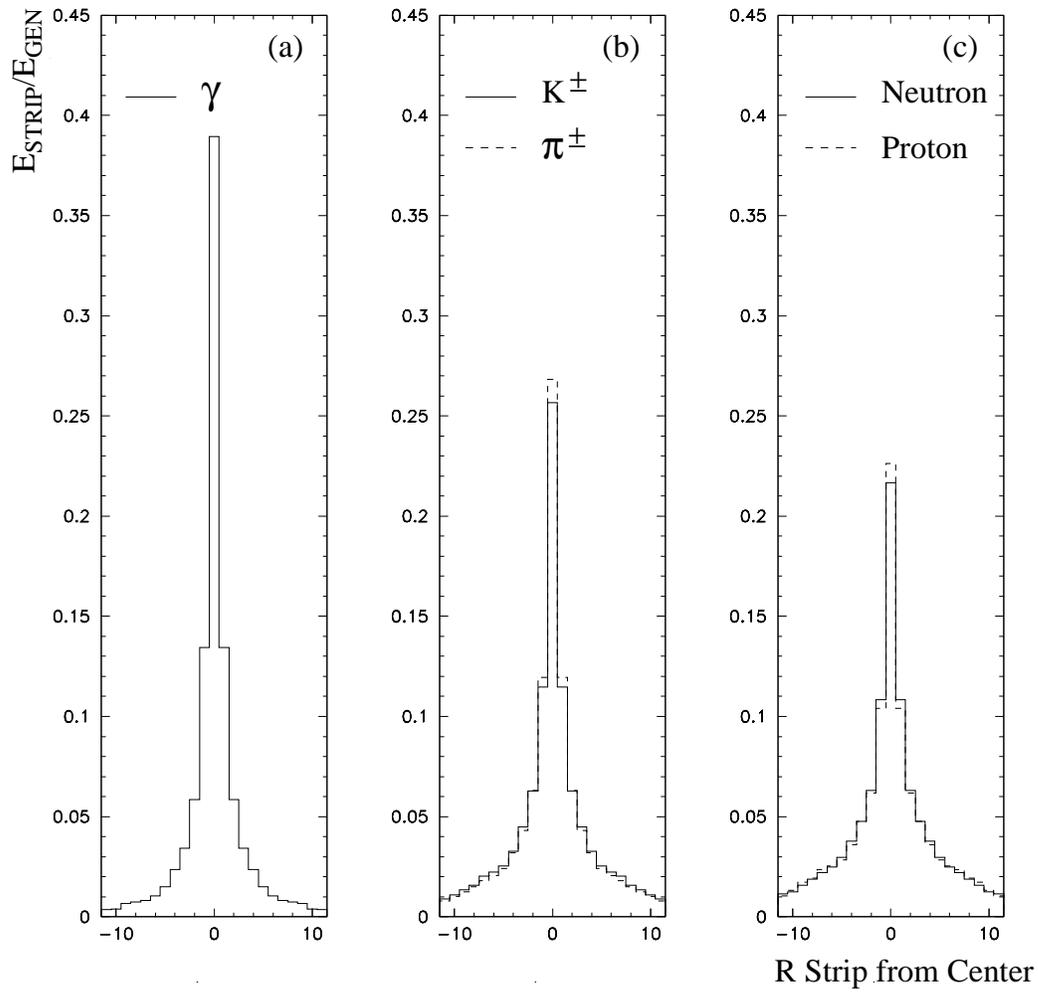}
\caption{Fractional energy deposition on the R strips of the EMLAC 
near the shower center for simulated incident 20~GeV (a) photons, (b) mesons, 
and (c) baryons.}
\label{shape}
\end{figure}

   The trigger simulation depends upon the modeling of the energy
deposited in the active volume of the EMLAC by incident particles.
In order to generate a substantial number of events in a timely manner,
the showers generated in the EMLAC by incident particles were parametrized.
After accounting for the energy loss in the inactive material in front
of the calorimeter (such as the cryostat wall),
the (calibrated) energy response of the EMLAC to incident photons and electrons
was parametrized by an energy resolution function, 
$\sigma_E = \sqrt{0.22^2 + 0.16^2 E + (0.01 E)^2}$,
where $E$ is the energy in GeV.
The response of the EMLAC to incident hadrons was investigated 
using the {\sc geant} software package\cite{geant}. 
Individual hadrons at a fixed energy of 20~GeV were generated 
in the target region and propagated through the full detector simulation 
(resulting in full showers in the EMLAC).
The simulated calibrated energy response relative to the incident hadron 
energy was evaluated for various incident hadron types and the resulting
distributions are shown in Fig.\ \ref{eop}(b) and (c).
[For comparison, the corresponding distribution for incident 20~GeV photons 
is shown in Fig.\ \ref{eop}(a).] 
These distributions were used to
parametrize the response of the EMLAC to incident hadrons. 
Since these distributions were generated based upon the full shower
response of the EMLAC, they already include the effects of the intrinsic 
energy resolution of the EMLAC.
The shapes of these distributions were found to be relatively insensitive to 
variations in the incident energy for $E~>~\approx$6~GeV, and hadrons 
below this energy were generally not detected. 
The transverse and longitudinal development of showers were 
also parametrized based upon full shower {\sc geant} studies.
In the transverse direction, a radial shower profile was used
to generate the simulated energy deposition on the R strips near the 
centroid of the shower (see Fig.\ \ref{shape}).
The longitudinal shower development for incident photons and hadrons was
parametrized as the ratio of the energy in the 
front section of the EMLAC with respect to the total energy 
deposited in the EMLAC\cite{blusk}. 
These parameterizations of the response of the EMLAC to incident
photons and hadrons were used to provide an efficient simulation of the
distribution of energy deposited in the EMLAC for each of the incident 
particles generated by {\sc pythia} which impinged on the active region of
the EMLAC.
In this manner, the total energy deposited on each R strip was
estimated as the scalar sum of the energy depositions of the
individual particles in that given event.

   The trigger $p_T$ detected by each R strip was evaluated by weighting 
the energy detected on that strip by the appropriate measured strip trigger 
weight, which increased as $\approx\sin{\theta_i}$. (The strip trigger weights
were measured and evaluated independently for each trigger type used in this
analysis.) From the $p_T$ in the strips, LOCAL and GLOBAL $p_T$ sums 
were calculated analogously to the online trigger method. 
These $p_T$ sums and the measured
trigger efficiency curves for each trigger were used to evaluate the
event trigger probabilities for each trigger type for each event.
The simulated event was either accepted or rejected 
based upon these trigger probabilities. 

\begin{figure}
\epsfxsize=6.5 truein
\epsffile[0 72 612 720]{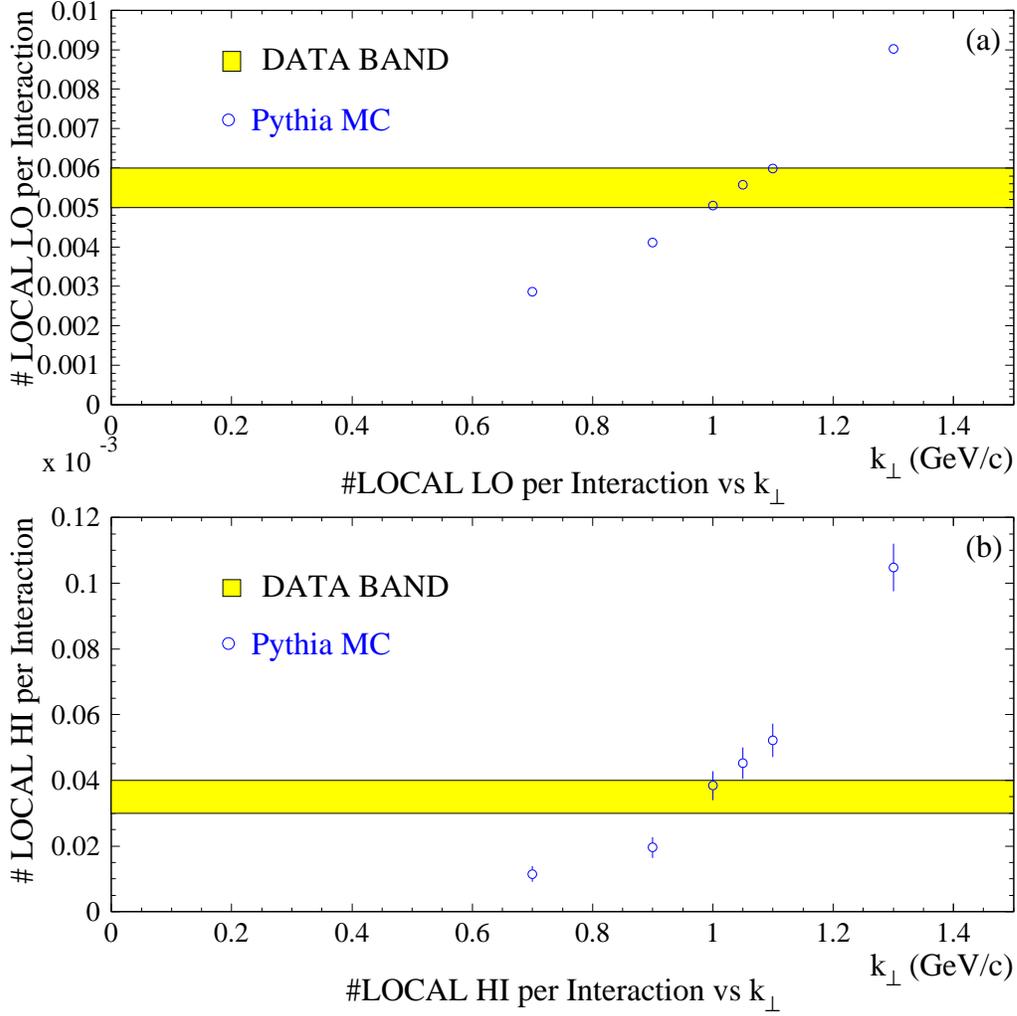}
\caption{The fraction of interactions in which (a) the LOCAL LO requirement was 
satisfied, and (b) the LOCAL HI requirement was satisfied, as a function of the 
\pythiakt\  parameter of the {\sc pythia} MC simulation. The shaded band across 
the plot shows the corresponding rates as measured in the low bias data. 
The width of the band is an estimate of the uncertainty in the rates measured 
in the low bias data sample.}
\label{ktcomp}
\end{figure}

  This trigger simulation was tested by comparing the fraction
of interactions in which the LOCAL LO 
and LOCAL HI requirements were satisfied in the Monte Carlo simulation
and the low bias data. The interactions in the low bias data
were required to have a reconstructed vertex in the target region
and events containing beam halo muons were excluded.
Several MC event samples were generated using the minimum
bias event generator and varying the \pythiakt\  parameter
from 0.7~GeV/$c$ to 1.3~GeV/$c$. For each of these samples, we measured the
rate at which the LOCAL LO and LOCAL HI signals were generated.
Figure \ref{ktcomp}(a) shows
the rate of LOCAL LO signal generation in the data sample
compared to the corresponding rate determined from
the MC samples. Figure \ref{ktcomp}(b) shows a similar comparison
for the LOCAL HI signal.
The open circles represent the MC results for the various \pythiakt\ 
choices, and the shaded band represents the data and its
associated uncertainty.
Both the LOCAL LO and LOCAL HI signal rates
are reproduced by the {\sc pythia} MC simulation using a 
\pythiakt\  value of $\approx$1~GeV/$c$. The 
relative rates between the SINGLE LOCAL HI, LOCAL$\otimes$GLOBAL HI, 
and TWO-GAMMA
triggers were also consistent between the MC and data samples. These 
observations indicate that the tuned MC simulation provides reasonable 
estimates of the rates at which high energy interactions 
satisfy the various triggers that contributed to this analysis.

\subsection{Trigger efficiency for charm events}

\begin{figure}
\epsfxsize=6.5 truein
\epsffile[0 72 612 720]{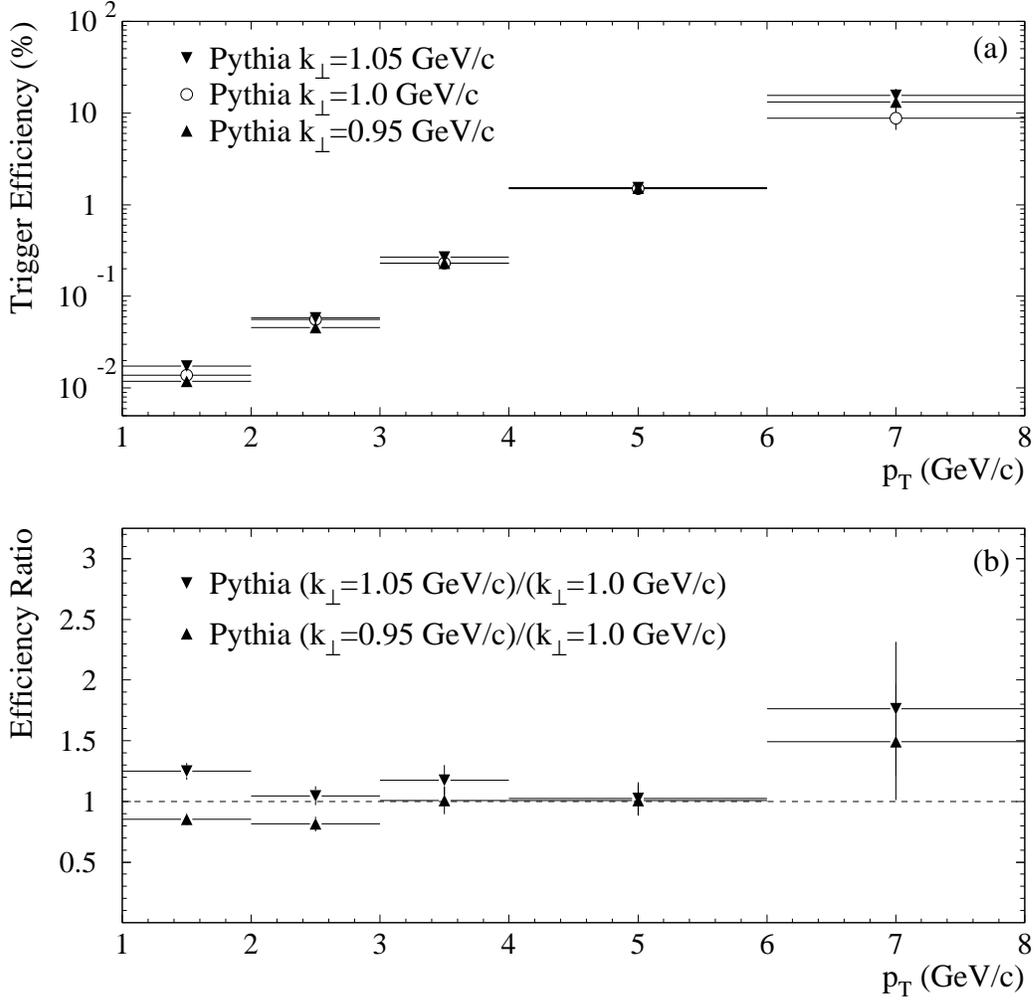}
\caption{(a) The trigger efficiency as a function of the $D^{\pm}$
$p_T$ for three values of \pythiakt; \pythiakt=0.95~GeV/$c$,  
\pythiakt=1~GeV/$c$, and \pythiakt=1.05~GeV/$c$, and (b) the ratios
of the trigger efficiencies evaluated with 
\pythiakt=1.05~GeV/$c$ compared to the central value of
\pythiakt=1.0~GeV/$c$ and \pythiakt=0.95~GeV/$c$ compared
to the central \pythiakt=1.0~GeV/$c$ value. The efficiencies shown
are averaged over the region $x_F>-0.2$.}
\label{kt-dependence}
\end{figure}

\begin{figure}
\epsfxsize=6.5 truein
\epsffile[0 72 612 720]{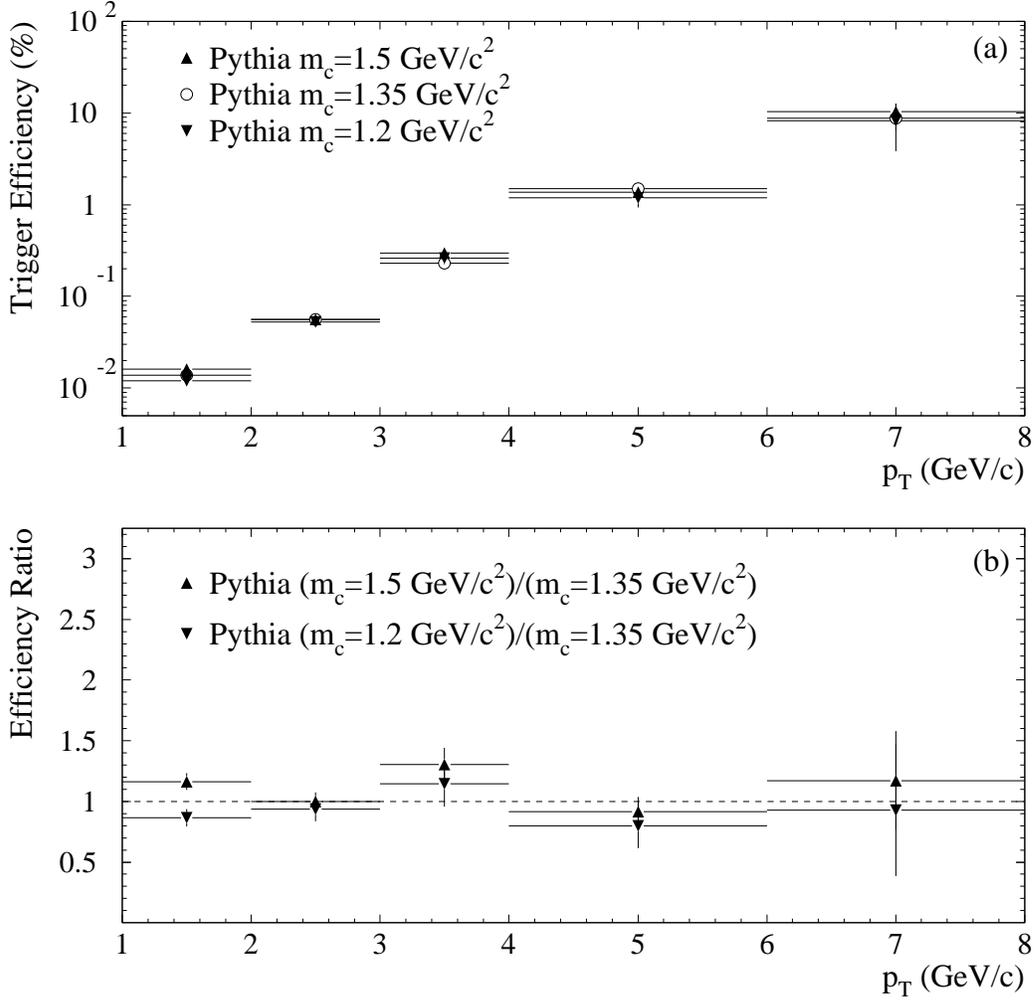}
\caption{(a) The trigger efficiency as a function of the $D^{\pm}$
$p_T$ for three values of the charm quark mass:
$m_c=1.5$~GeV/$c^2$, $m_c=1.35$~GeV/$c^2$ and $m_c=1.2$~GeV/$c^2$,
and (b) the ratio of trigger efficiencies evaluated with the larger 
$m_c=1.5$~GeV/$c^2$ compared to the central $m_c=1.35$~GeV/$c^2$ mass value
and the smaller $m_c=1.2$~GeV/$c^2$ compared to the central 
$m_c=1.35$~GeV/$c^2$ mass value. The default value in the {\sc pythia} 
MC simulation is 1.35~GeV/$c^2$.
The efficiencies shown are averaged over the region $x_F>-0.2$.}
\label{mass-dependence}
\end{figure}

  Since the tuned {\sc pythia} MC simulation reproduced the kinematic 
distributions of final state hadrons, as well as the observed trigger 
rates in the data, we used this simulation to evaluate 
the trigger efficiency for charm events. The trigger efficiency
is determined by calculating the probability that an event containing a 
$D^{\pm}$ meson (which decays to $K\pi\pi$) will satisfy
the SINGLE LOCAL HI, LOCAL$\otimes$GLOBAL HI, or TWO-GAMMA trigger.
(We compared the integrated $D^{\pm}$ cross sections determined 
using each trigger type individually, and found them to be consistent
within uncertainties.) The average trigger efficiency is evaluated
as a function of the $p_T$ of the charged $D$ meson that decayed 
to the $K\pi\pi$ final state. Figure \ref{kt-dependence}(a) shows 
the resulting trigger efficiency for the central 
\pythiakt\  value of 1~GeV/$c$, as well as for other reasonable choices 
of \pythiakt\  values
based upon Fig.\ \ref{ktcomp}. In Fig.\ \ref{kt-dependence}(b), the
ratios of the trigger efficiencies for the larger (and smaller) 
\pythiakt\  values with respect to the central value are presented.
The $\pm 15\%$ systematic uncertainty associated with this choice was 
determined based on the fractional difference 
between the mean value of the efficiencies determined via the larger
and smaller \pythiakt\  choices. It is also plausible that the trigger
efficiency will be sensitive to the value of the charm quark mass
that is used in the MC simulation.
Figure \ref{mass-dependence}(a) shows the trigger efficiency versus the
transverse momentum of the $D^{\pm}$ for three choices of $m_c$
(1.2, 1.35, and 1.5~GeV/$c^2$). 
Figure \ref{mass-dependence}(b)
shows the ratio of trigger efficiencies for the larger and smaller
charm quark mass values with respect to the central value.
The $\pm 10\%$ uncertainty associated with the choice of $m_c$=1.35~GeV/$c^2$
was determined based upon the fractional change in 
the mean value of the trigger efficiency when $m_c$ was varied from
1.2~GeV/$c^2$ to 1.5~GeV/$c^2$. 

\subsection{Tracking simulation}
  
\begin{figure}
\epsfxsize=6.5 truein
\epsffile[0 72 612 720]{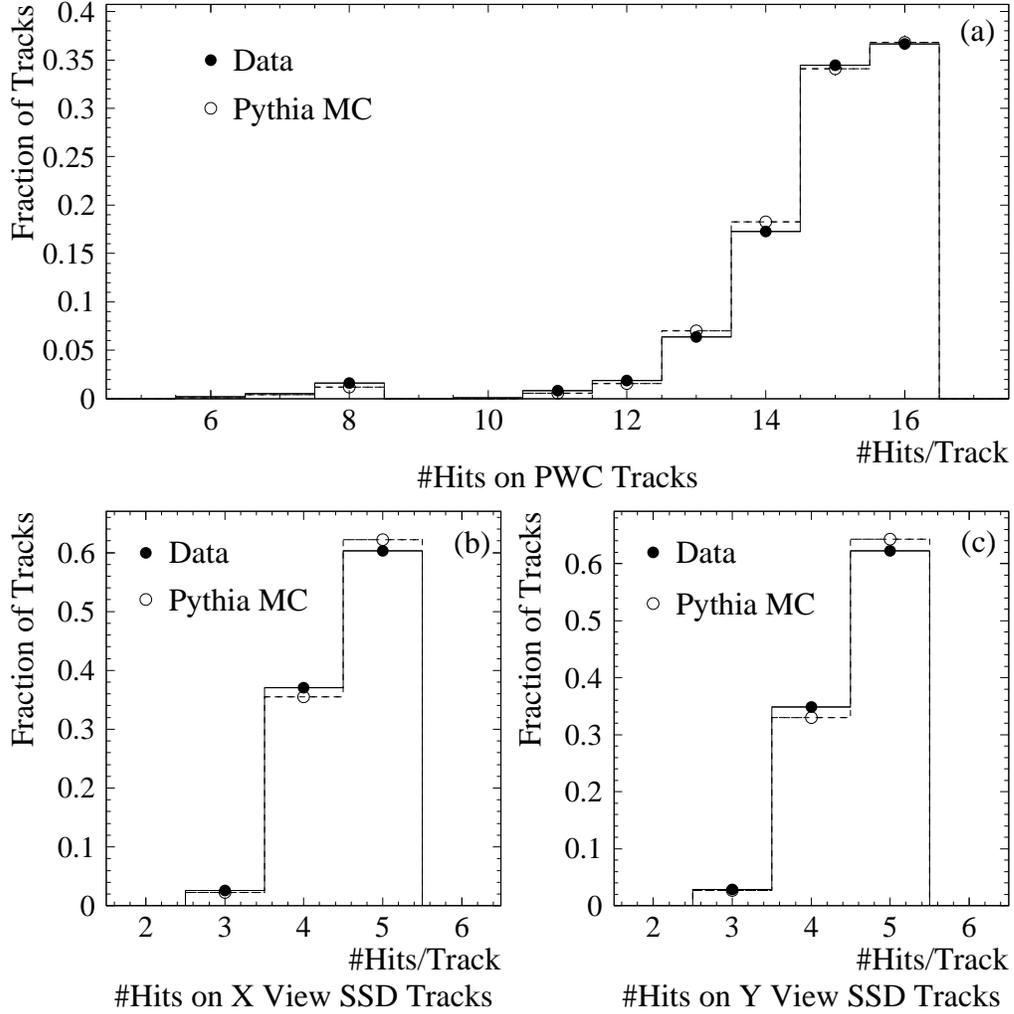}
\caption{The normalized distribution of the number of hits on 
reconstructed tracks in (a) the PWC system, (b) the SSD $X$ view, 
and (c) the SSD $Y$ view. The filled circles
represent the distribution measured in the data while the open
circles are the corresponding results from the MC simulation.}
\label{hit-trk}
\end{figure}

\begin{figure}
\epsfxsize=6.5 truein
\epsffile[0 72 612 720]{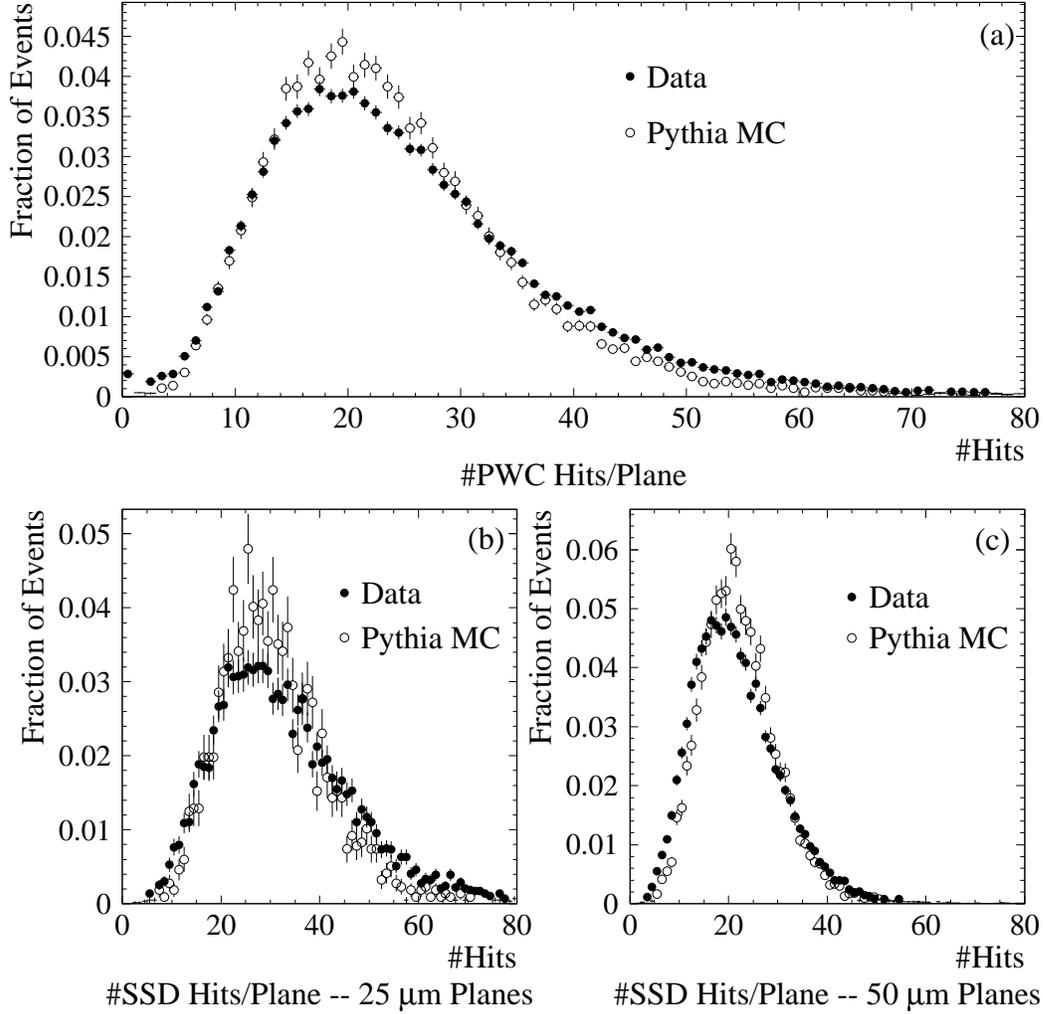}
\caption{The normalized distribution of the total number of hits 
per plane in (a) the PWC's,
(b) the vertex SSD planes that have 25~$\mu$m regions, and
(c) the other vertex SSD planes. The filled circles
represent the distributions measured in the data and the open circles 
are the corresponding results from the MC simulation.}
\label{hits}
\end{figure}

\begin{figure}
\epsfxsize=6.5 truein
\epsffile[0 72 612 720]{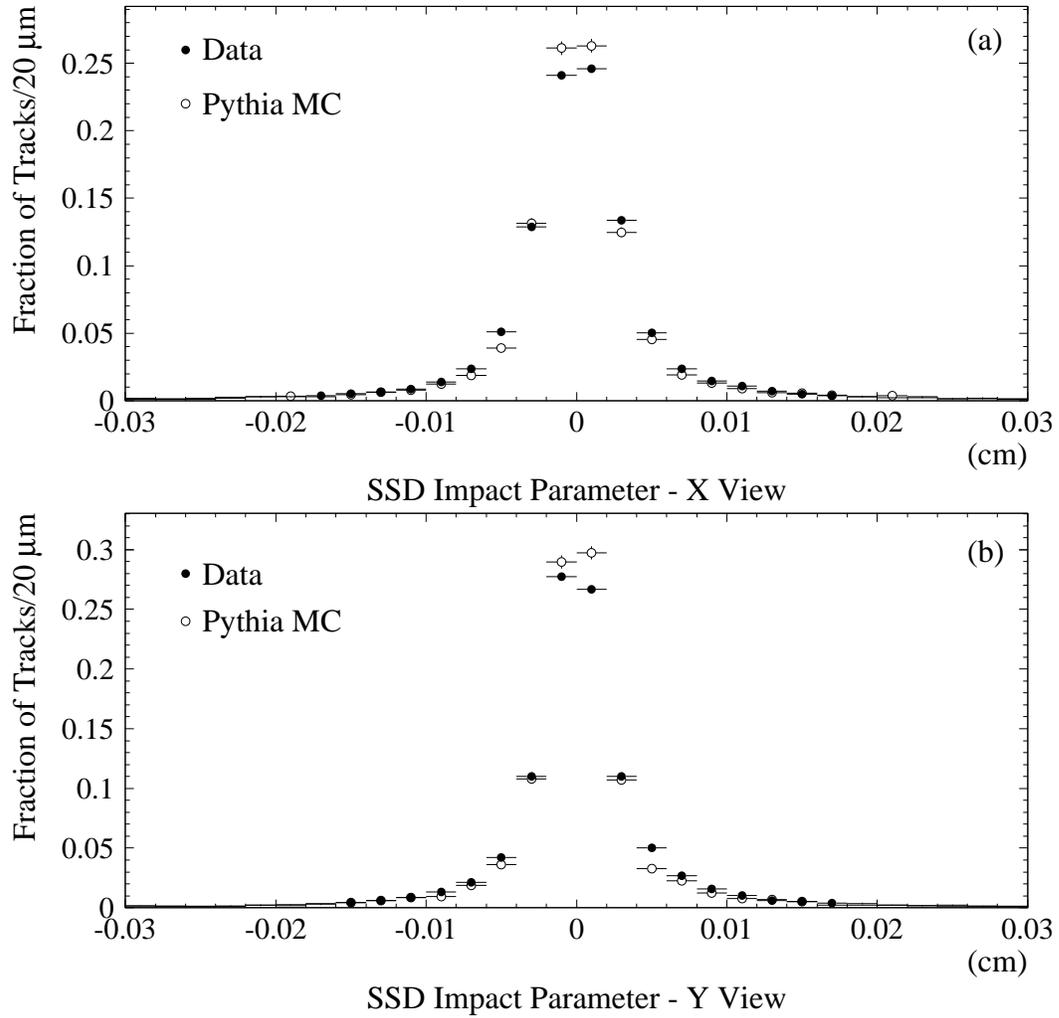}
\caption{Impact parameter distribution of SSD tracks to
the primary vertex in the (a) $X$ and (b) $Y$ views. The filled circles
represent the distributions measured in the data and the open circles 
are the corresponding results from the MC simulation.}
\label{ssd-imp}
\end{figure}

    In addition to correcting for losses due to the trigger,
we must also evaluate the efficiency of
reconstructing the decay vertex once the event has triggered the 
apparatus. All of the tracking detectors were modeled within the
framework of {\sc geant} to estimate the losses due to the 
geometrical acceptance of the tracking system and detector
performance. Various detector effects, such as efficiency, resolution
and noise, were evaluated in the data and then incorporated into the MC 
simulation. 
Hit efficiencies are modeled as a function of transverse position
in the tracking detectors.
Figure \ref{hit-trk} shows the number of hits on the reconstructed 
tracks in the PWC system, and the $X$ and $Y$ views of the SSD system
for both the MC and data samples. The 
consistency between these distributions indicates that the tracking detector 
efficiencies are well modelled. We also tuned the
MC simulation to reproduce the hit multiplicities in the charged tracking
detectors. Figure \ref{hits} shows the MC and data
distributions for the average hit multiplicity in the PWC planes,
and the 25~$\mu$m and 50~$\mu$m pitch SSD planes, respectively.
Both the effects of delta rays and noise hits have been 
included in this simulation\cite{chung}. The mean hit multiplicity
from the data and the MC simulation are similar; however, the data 
distributions are slightly broader.  
The impact of increasing the hit multiplicity assumed in the MC simulation 
on the charged $D$ meson reconstruction efficiency is discussed in 
the next section.
Finally, we investigate
how well the resolution of the detector is simulated.
The impact parameter distribution provides
a measure of the angular precision of the SSD tracks,
and this is pertinent to separating heavy quark decay vertices from the
primary vertex. Figure \ref{ssd-imp} shows the MC and data 
impact parameter distributions of all charged tracks relative to the 
primary vertex for the $X$ and $Y$ track segments of the SSD system. 
The MC distribution is $\approx$5\% narrower than the corresponding data
distribution. The
effect of this difference on the estimated $D^{\pm}$ reconstruction
efficiency will be addressed in the next section.

\begin{figure}
\epsfxsize=6.5 truein
\epsffile[0 72 612 720]{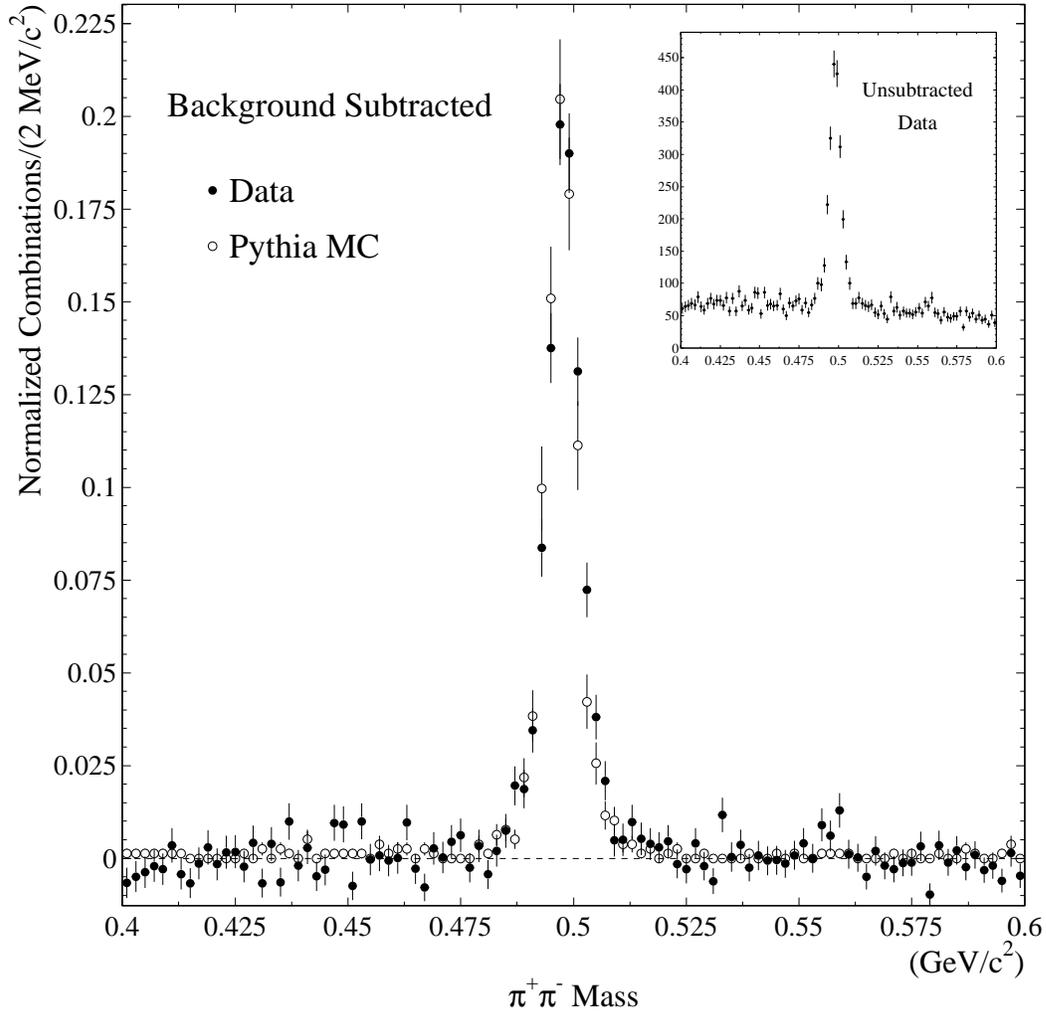}
\caption{Background subtracted $\pi^+\pi^-$ invariant mass distribution 
for vees reconstructed in the MC and data samples. The filled circles
are the data and the open circles are the 
corresponding results from the MC simulation. The inset shows the unsubtracted 
$\pi^+\pi^-$ invariant mass spectrum for the data events.}
\label{k0-signal}
\end{figure}

\begin{figure}
\epsfxsize=6.5 truein
\epsffile[0 72 612 720]{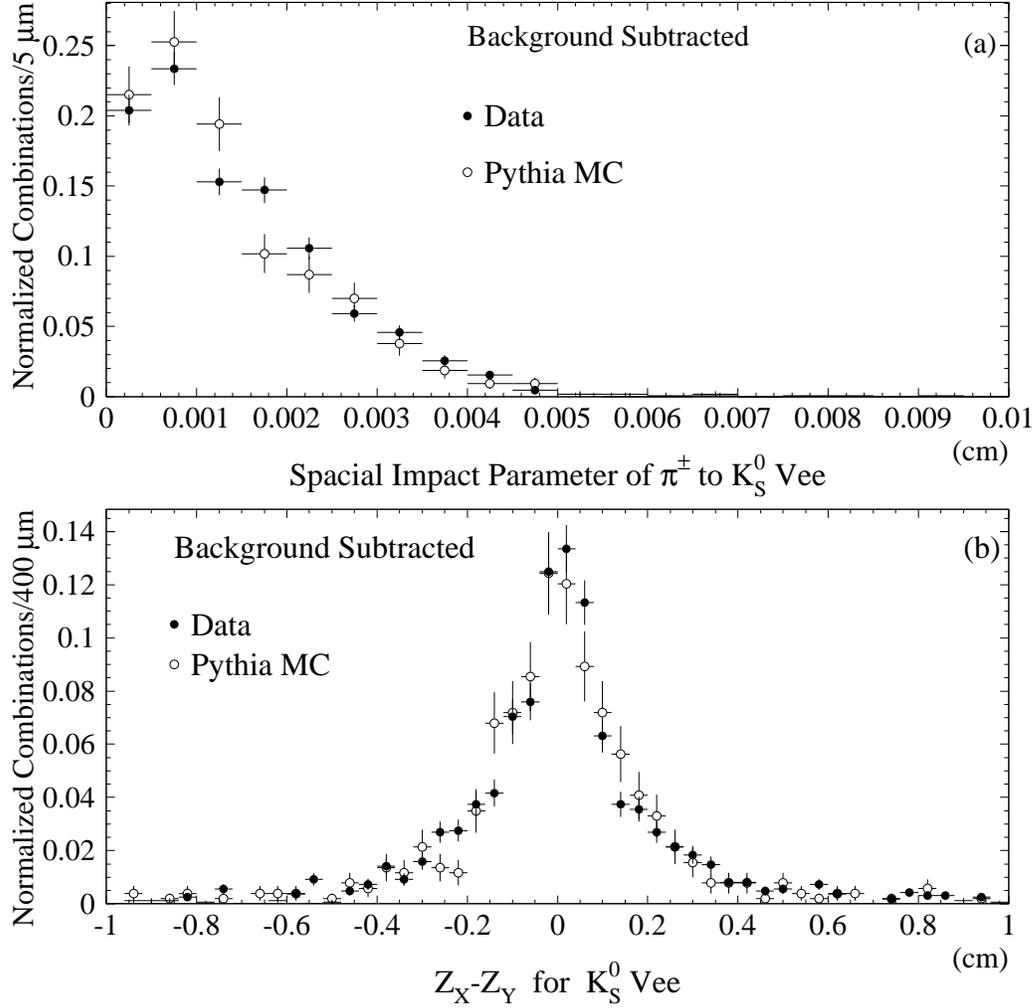}
\caption{(a) Spatial impact parameter distributions of the pion tracks to the 
$K^0_S$ vee location, and (b) the difference in the reconstructed $Z$ location
of the vee between the $X$ and $Y$ views. Both distributions
are background subtracted. The filled circles
are the distributions from the data and the open circles 
are the  corresponding results from the MC simulation.}
\label{k0-rec}
\end{figure}

  The decay of $K^0_S$ into $\pi^+\pi^-$ provides an 
opportunity to compare high statistics MC and data distributions
relevant to secondary vertex finding. A large sample of
$K^0_S$ mesons were reconstructed in the data using
displaced two track secondary vertices.
A sample of full events enriched with
$K^0_S$ mesons was generated using the minimum bias event
generator from {\sc pythia}, and reconstructed using the
secondary vertex algorithm described in this paper.
The background subtracted $K^0_S$ signals are shown in
Fig.\ \ref{k0-signal}. The MC and data $K^0_S$ mass
resolutions are in good agreement.
Figure \ref{k0-rec}(a) shows the 3D impact parameter distributions
of the $\pi^+$ and $\pi^-$ tracks (from the $K^0_S$
signal region) to the vee location. Figure \ref{k0-rec}(b) shows the
difference in the reconstructed $Z$ coordinate of the vee
as determined in the $X$ and $Y$ views independently. In both cases,
the MC simulation reproduces the distributions observed in the data.
We have compared the distributions of other analysis variables 
as well and found similar agreement\cite{blusk}.

\subsection{Charged $D$ meson reconstruction efficiency}

\begin{figure}
\epsfxsize=6.5 truein
\epsffile[0 72 612 720]{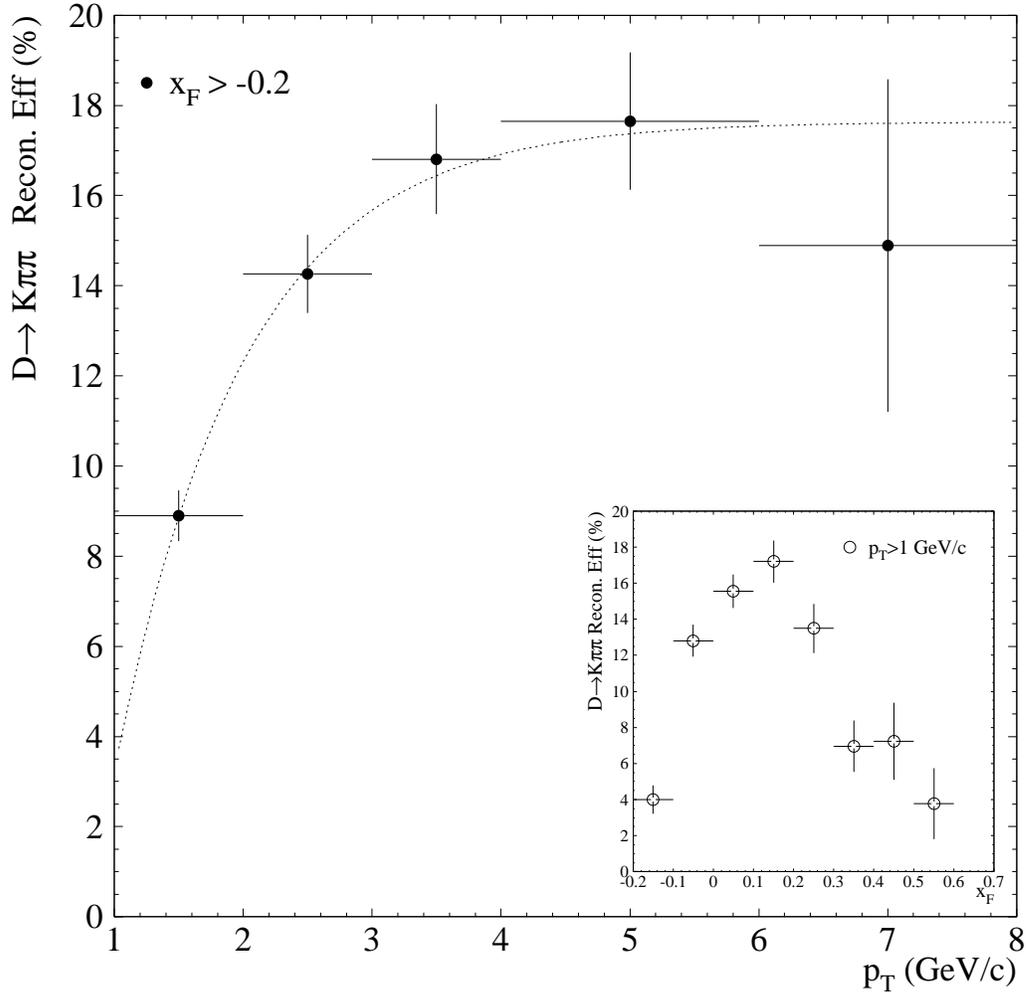}
\caption{$D^{\pm}\to K^{\mp}\pi^{\pm}\pi^{\pm}$ reconstruction efficiency 
as a function of the reconstructed $D^{\pm}$ transverse momentum for 
events that satisfied at least one of the high $p_T$ triggers used in this 
analysis.
The curve is a parametrization of the efficiency as a function of $p_T$.
The inset shows the $D^{\pm}\to K^{\mp}\pi^{\pm}\pi^{\pm}$ reconstruction 
efficiency as a function of $x_F$ for $D^{\pm}$ mesons with $p_T>1$~GeV/$c$ 
for triggered events.  Error bars reflect statistical uncertainties only.}
\label{eff_paper}
\end{figure}

  The $D^{\pm}$ reconstruction 
efficiency, which includes acceptance losses, detector 
effects, and analysis requirements, is evaluated using the sample
of {\sc pythia} charm events generated with \pythiakt=1.0~GeV/$c$ that
satisfy at least one of the high $p_T$ triggers. 
Figure \ref{eff_paper} shows the $D^{\pm}$ reconstruction 
efficiency as a function of the reconstructed $D^{\pm}$ transverse 
momentum.  The average reconstruction efficiency for each $p_T$ bin 
is defined as the value of the parametrization shown in 
Fig.\ \ref{eff_paper} evaluated at the center of the bin. 
The reconstruction efficiency increases from $\approx$9\% in
the lowest $p_T$ bin (1 to 2~GeV/$c$) to $\approx$17\% in our highest
$p_T$ bin (6 to 8~GeV/$c$).  The inset in Fig.\ \ref{eff_paper}
illustrates the $D^{\pm}$ reconstruction efficiency versus $x_F$ for
$D^{\pm}$ mesons with $p_T>1$~GeV/$c$.

   To estimate the systematic uncertainty in the
reconstruction efficiency, two additional versions of the
MC simulation were
evaluated. One version included more noise hits than the default
version, and the other had reduced hit multiplicity. The
extreme values were determined based upon variations observed
in the data at different times during the data taking period. 
The variations in hit multiplicity were typically less than 
10\% of the central value. The primary causes of these 
variations in the average hit multiplicity are 
variations in the beam intensity and detector performance.
Increasing the number of noise hits results in a degradation
of the track and vertex resolution which reduces the $D^{\pm}$
reconstruction efficiency. Reducing the hit multiplicity 
has the opposite effect.
We compared the impact parameter distributions
of charged tracks to the primary vertex for these two MC versions
and found that the higher hit multiplicity version more accurately 
reproduced the data result. However, the hit and track
multiplicity in this version of the MC was larger than that 
observed in the data. The sensitivity of the $D^{\pm}$ reconstruction
efficiency to the choice of either matching the track multiplicity distribution
or the impact parameter distribution from the MC simulation and the data
reflects an uncertainty in the reported reconstruction efficiency.
The 10\% systematic uncertainty in the $D^{\pm}$ reconstruction efficiency
is based on the relative difference between the results from 
these two versions of the MC simulation. 

\section{RESULTS}
   
    The observed $D\to K\pi\pi$
signals and the efficiencies for triggering on
and reconstructing these events are tabulated
in Tables \ref{signal-allxf-table} and \ref{signal-posxf-table}
for charged $D$ mesons in the kinematic ranges $x_F>-0.2$ and 
$x_F>0.0$, respectively. Using these
numbers, the cross section in each $p_T$ bin is given by 
$\frac{N(p_T)}{\epsilon(p_T)LB}$, where
$N(p_T)$ is the number of mass combinations above background
in the given $p_T$ bin,
$\epsilon(p_T)$ is the efficiency for reconstructing those events,
$L$ is the integrated luminosity, and $B$ is the branching ratio
for $D\to K\pi\pi$, which is 9.1$\pm$0.6\%\cite{pdg}.
The integrated luminosity per nucleon is 7.8$\pm$0.8 events/pb
for this data sample, including target transverse fiducial cuts 
and corrections for beam absorption. 
In combining the signals from the beryllium and copper targets, nuclear
effects were assumed to be negligible for the hadroproduction of open charm.
The dependences of cross sections per nucleus on atomic mass are often
parametrized as $\sigma_{0}$$A^{\alpha}$, where $A$ is the atomic mass 
of the target nucleus.  Recent measurements of $D$
meson production yield values of $\alpha$ consistent with 
one\cite{dameri,alpha,nu-e769}.  No significant $p_T$ (or $x_F$) dependence 
of $\alpha$ is observed\cite{nu-e769}.
The differential and integrated 
$D^{\pm}$ cross sections are discussed in the following 
subsections. 

\subsection{Differential cross sections}

\begin{figure}
\epsfxsize=6.5 truein
\epsffile[0 72 612 720]{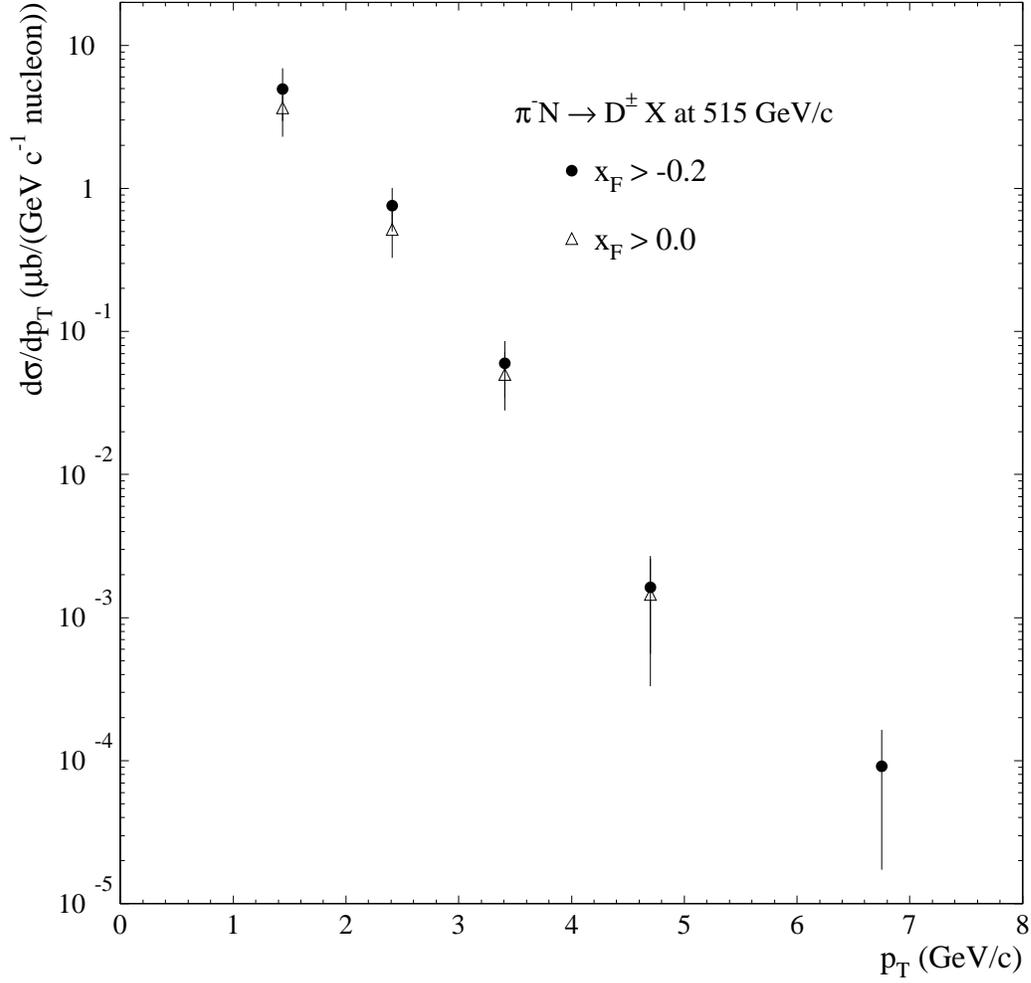}
\caption{Differential cross section per nucleon for
$D^{\pm}$ production in 515~GeV/$c$ $\pi^-$-nucleon
collisions as a function of the $p_T$ of the $D$ meson. The filled circles
represent data integrated over the region $x_F>-0.2$ and the triangles
are for data integrated over $x_F>0$. The vertical
error bars represent the statistical and systematic uncertainties added
in quadrature.}
\label{dfxs-pt}
\end{figure}

  The $D^{\pm}$ differential cross sections per nucleon integrated over 
the ranges $x_F>-0.2$ and $x_F>0$ are presented in 
Table \ref{xs-table}\cite{xserr} and
displayed in Fig.\ \ref{dfxs-pt}.
Due to the steeply falling spectra and the
large widths of the $p_T$ bins, the data points are plotted at
$p_T$ values, $p_T^{lw}$, that correspond to the average values 
of the cross section in the appropriate bins as determined
from the {\sc pythia} MC $p_T$ spectra\cite{laff}. 
The uncertainty in the cross 
section is obtained from the quadrature sum of the statistical and
systematic uncertainties. 

\begin{figure}
\epsfxsize=6.5 truein
\epsffile[0 72 612 720]{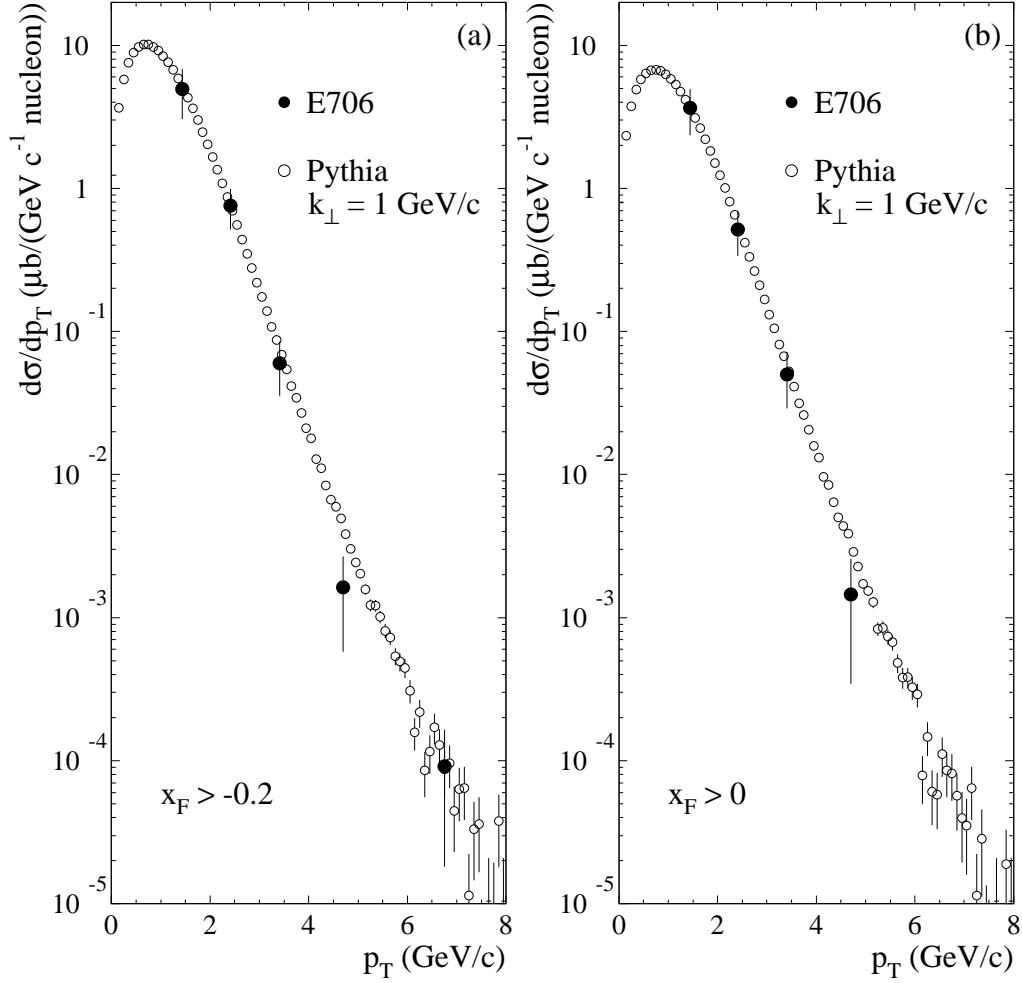}
\caption{Comparison of the $D^{\pm}$ differential cross 
section per nucleon measured by experiment E706 with the corresponding result
from the {\sc pythia} simulation MC for the kinematic ranges (a) $x_F>-0.2$, 
and (b) $x_F>0$. These {\sc pythia} results are normalized to match our 
measured cross section integrated over $x_F>-0.2$ and $1<p_T<8$~GeV/$c$. 
The error bars for the data represent statistical and systematic uncertainties
added in quadrature excluding luminosity and branching ratio contributions.}
\label{e706-pythia}
\end{figure}

   Figure \ref{e706-pythia} shows our measured $D^{\pm}$ cross sections 
and the results of the {\sc pythia} MC simulation.  Common systematic 
uncertainties in the luminosity and branching ratio have been excluded
from the uncertainties shown in this figure. 
The shapes of the $D^{\pm}$ spectra generated by the {\sc pythia} MC simulation 
are consistent with our measured spectra. 
These {\sc pythia} results were normalized to our measured cross section 
integrated over $x_F>-0.2$ and $1<p_T<8$~GeV/$c$.  The consistency    
between the data and the corresponding {\sc pythia} result integrated 
over the range $x_F>0$ [shown in Fig.\ \ref{e706-pythia}(b)] indicates
that {\sc pythia} adequately models the fraction of the cross section 
in the range $x_F>0$ relative to the fraction in the range $x_F>-0.2$.

\begin{figure}
\epsfxsize=6.5 truein
\epsffile[0 72 612 720]{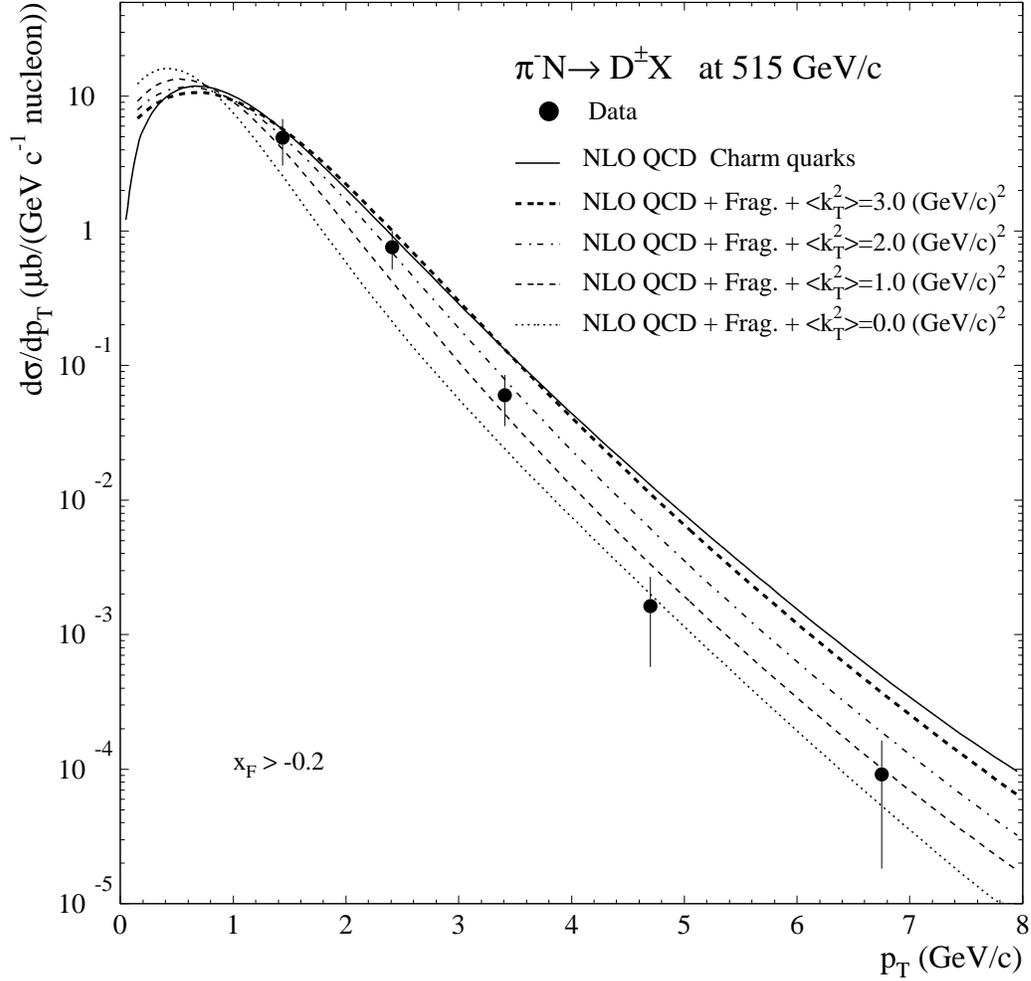}
\caption{Comparison of the $D^{\pm}$ differential cross section per nucleon 
measured by experiment E706 to the results of NLO pQCD calculations. The pure 
NLO pQCD unfragmented result, and the NLO Peterson {\it et al.} fragmented 
spectra supplemented with average intrinsic transverse momenta squared 
(\NLOkt2) of 0.0~(GeV/$c$)$^2$, 1.0~(GeV/$c$)$^2$, 2.0~(GeV/$c$)$^2$, 
and 3.0~(GeV/$c$)$^2$ are 
shown in this figure. The charm quark mass employed in this
calculation is 1.5~GeV/$c^2$.  Error bars represent statistical and 
systematic uncertainties added in quadrature excluding luminosity and 
branching ratio contributions.}
\label{e706-nlo-pt-allxf}
\end{figure}

  Figure \ref{e706-nlo-pt-allxf} shows 
the differential charged $D$ cross section compared to results of NLO pQCD 
calculations\cite{mangano_p}.  The renormalization and factorization 
scales employed in the calculation of these differential distributions 
are as follows: $\mu_R=\mu_0$ and $\mu_F=2\mu_0$, where 
$\mu_0=\sqrt{m_c^2+\frac{1}{2}(p_T^2+\overline{p}_T^2)}$ and
$p_T$ and $\overline{p}_T$ are the transverse momenta of the charm and 
anti-charm quarks.
The solid curve shows the NLO pQCD result 
for charm quark production.  The dotted curve illustrates the NLO pQCD result
including the effects of Peterson {\it et al.} fragmentation
(with $\epsilon_c$=0.06).
The other broken curves show the Peterson {\it et al.} fragmented 
NLO pQCD results for charm production supplemented with
various values of \NLOkt2. The pQCD cross sections
shown in Fig.\ \ref{e706-nlo-pt-allxf} have been normalized to
our extrapolated integrated $D^{\pm}$ cross section (see next section).
The results are integrated over the range $x_F>-0.2$. 
(The corresponding comparisons for results integrated over
the range $x_F>0$ do not appear substantially different.) 
The common systematic 
uncertainties in the luminosity and branching ratio have been excluded
from the uncertainties shown in this figure.  
The shapes of the 
theoretical $p_T$ spectra for unfragmented charm quark production
are consistent with data on the hadroproduction 
of $D$ mesons from experiments E769\cite{e769-new2} and WA82\cite{mangano5}.
E653 reported that the shape of the unfragmented NLO pQCD transverse
momentum distribution was ``somewhat harder'' than the observed 
distribution\cite{e653}.
Our data probe a larger $p_T$ range, and the unfragmented theoretical 
$p_T$ spectrum is clearly harder than the spectrum observed in our data.
Introducing Peterson {\it et al.} fragmentation 
into the calculation results in a softer $p_T$ spectrum whose shape better
matches the shape observed.  It is clear that the NLO pQCD
spectrum generated using $m_c$=1.5~GeV/$c^2$ and supplemented 
with \NLOkt2=3~(GeV/$c$)$^2$ is harder than our data,
while the corresponding $p_T$ spectra generated with \NLOkt2\ values 
between 1 and 2~(GeV/$c$)$^2$ are similar to what we observe.
As noted in the Introduction, the $D$ meson $p_T$ spectra reported by
experiments E769 and WA82 can also be described by the fragmented NLO pQCD 
results supplemented with a \NLOkt2=2~(GeV/$c$)$^2$\cite{mangano5}.

\subsection{Integrated cross section}

   Since the shapes of the E769 and WA82 $D$ meson $p_T$ spectra are 
consistent with the NLO pQCD calculations using Peterson {\it et al.} 
fragmentation supplemented with \NLOkt2=2~(GeV/$c$)$^2$,
we used that pQCD calculation to estimate the extrapolation factor 
necessary to account for the low $p_T$ portion ($p_T<1$~GeV/$c$) of the 
$D^{\pm}$ cross section. We find an extrapolation factor of 2.7.
The uncertainty in that factor is estimated to be 15\%
(which corresponds to the range of values obtained from the fragmented 
NLO calculation supplemented with \NLOkt2=1.5~(GeV/$c$)$^2$
and the unfragmented NLO pQCD calculation). An additional
factor of 1.07$\pm$0.03 accounts for the $D^{\pm}$ mesons produced 
with $x_F<-0.2$. This factor was evaluated using
the $x_F$ spectrum of $D^{\pm}$ mesons from the {\sc pythia} 
MC simulation. The uncertainty was based on half the difference between the
factor predicted for $D^+$ and $D^-$ mesons separately.

\begin{figure}
\epsfxsize=6.5 truein
\epsffile[0 72 612 720]{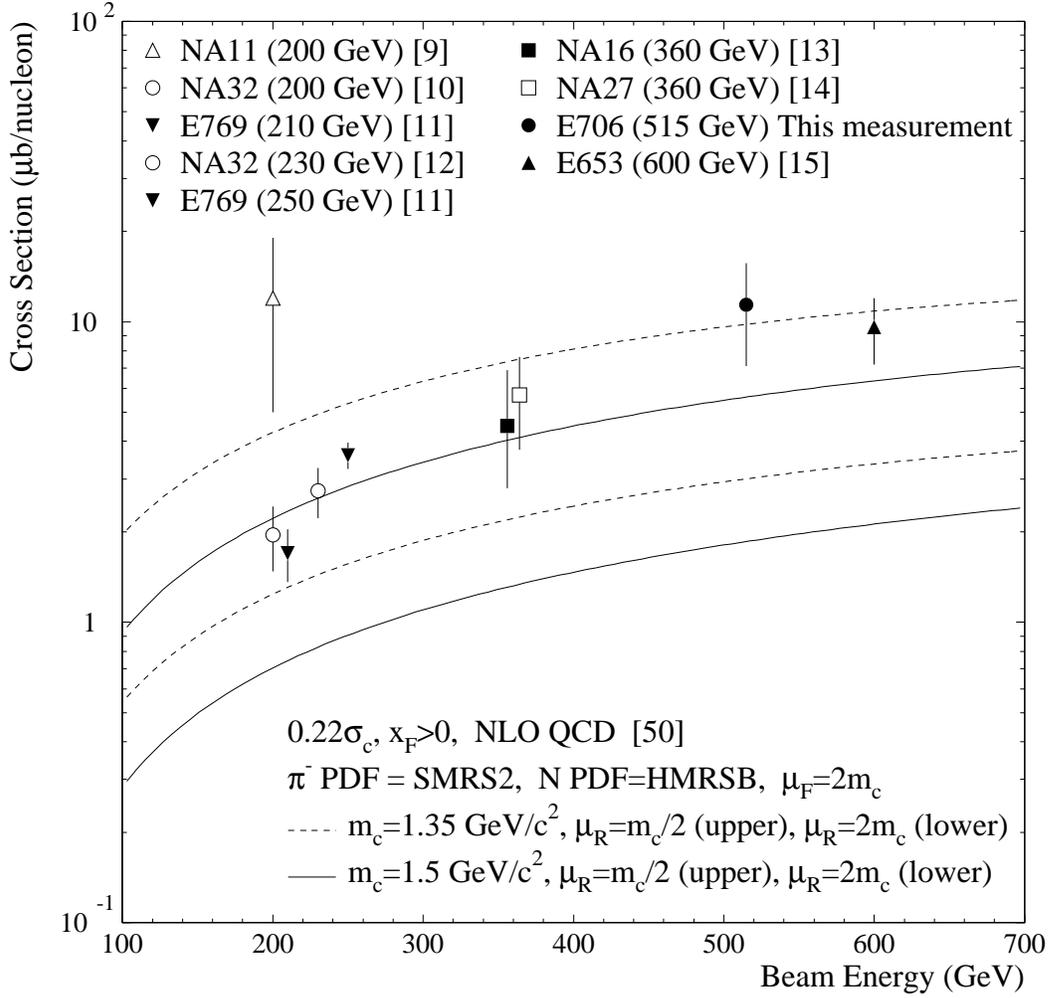}
\caption{The integrated cross section for inclusive $D^{\pm}$ production
with $x_F>0$ for incident $\pi^-$ as a function of beam energy.
Also shown are NLO pQCD calculations of the charm cross section in 
the region $x_F>0$ for two choices of the charm quark mass (1.5~GeV/$c^2$
and 1.35~GeV/$c^2$)  and two choices of the
renormalization scale ($\frac{1}{2}$$m_c$ and $2m_c$).
In each case, the theoretical cross section is calculated
with the factorization scale fixed at $\mu_F=2m_c$.
(Both the NA16 and NA27 measurements were at 360~GeV, but are plotted at 
356 and 364~GeV respectively, in order to clearly show the individual 
data points and their uncertainties.)} 
\label{dxs}
\end{figure}

Using these factors, the total
$D^{\pm}$ cross section is 
$16.6\pm 4.5(stat) \pm 4.8(syst) \ \mu$b
per nucleon. The systematic error estimate includes
uncertainties in the trigger and reconstruction efficiencies,
normalization, branching ratio, and the extrapolation.
The $D^{\pm}$ cross section for $x_F>0$ is
$11.4\pm 2.7(stat)\pm 3.3(syst) \ \mu$b per nucleon.
This result is compared  in Fig.\ \ref{dxs} with previous 
measurements \cite{na11,na32a,e769-new,na32,na16,na27,e653}
of the inclusive $D^{\pm}$ cross section 
in $\pi^-$ interactions at other beam energies.
Where appropriate, the results from other experiments have been adjusted 
to reflect current branching ratio values.  The error bars
represent the statistical and systematic uncertainties added in quadrature.
The figure also shows the results of NLO pQCD calculations of the 
charm cross section\cite{nason3,nason1,mangano_p}. 
The results have been adjusted by
a factor of 2 to account for the associated production of charm 
and are reduced by a factor of 1.6 to reflect the partial $x_F$ coverage
($x_F>0$)\cite{mangano4,xffac}. Since it is expected that the $D^{\pm}$
to charm fraction is nearly constant over this energy range\cite{mangano5}, 
the results are also multiplied by a factor of 0.22 
to account for the fragmentation process $c\to D^{\pm}$.
This factor was extracted from published measurements of the 
forward production charm hadron cross sections\cite{e769-new}.
The pQCD calculations use the SMRS2\cite{smrs2} and 
HMRSB\cite{hmrsb} parton distribution functions for the pion and 
nucleons, respectively.
The solid curves in the figure are the values of the calculated 
charm cross section generated using a charm quark mass of $m_c=1.5$~GeV/$c^2$
and renormalization scales of $2m_c$ and $\frac{1}{2}$$m_c$. To illustrate 
the sensitivity of the calculation to variations in the input parameters, 
the corresponding results are also displayed for a charm quark
mass of 1.35~GeV/$c^2$ (dashed curves).  The calculations are expected to
exhibit similar sensitivities to
other inputs, such as the factorization scale and 
parton distribution functions.
While these theoretical 
uncertainties in the normalization of the charm cross section are large 
at NLO, the energy dependence of the calculated
charged $D$ meson cross section is less sensitive to these uncertainties and
adequately describes the energy dependence 
suggested by the data points from the various experiments.

\section{CONCLUSIONS}

   We have analyzed a large sample of 515~GeV/$c$ $\pi^-$
interactions in copper and beryllium targets selected via a high 
transverse momentum electromagnetic 
shower trigger to study the hadronic production of high transverse
momentum charged $D$ mesons. Secondary vertices from
$D^{\pm}\to K^{\mp}\pi^{\pm}\pi^{\pm}$ decays were reconstructed
in a charged particle tracking system which included silicon strip detectors, 
a dipole magnet, proportional wire chambers, and straw tube drift chambers.
The data span the kinematic range 
$x_F>-0.2$ and $1<p_T<8$~GeV/$c$, a range which exceeds
previously reported measurements. 

   The measured differential
cross section is consistent with results from {\sc pythia}, provided 
{\sc pythia} is supplemented with a \pythiakt$\approx$1~GeV/$c$. 
Our measured charged $D$ meson $p_T$ spectrum
is not well described by the NLO pQCD calculations for charm quark
production that describe the lower $p_T$ results reported by 
E769\cite{e769-new2}.
Our data are
consistent with the charm $p_T$ spectra resulting from  
Peterson {\it et al.} fragmented NLO pQCD calculations supplemented 
with a \NLOkt2\ of 1 to 2~(GeV/$c$)$^2$. 
This observation is consistent with
previous reports on the comparison of the measured inclusive $p_T$ spectra of 
hadroproduced charm particles to fragmented NLO pQCD calculations 
supplemented with additional \NLOkt2\cite{mangano5}.

   An extrapolation based upon our measured spectrum
yields an integrated $D^{\pm}$ cross section of 
$11.4\pm 2.7(stat)\pm 3.3(syst) \ \mu$b 
per nucleon for $x_F>0.0$.  This value is consistent with the trend
observed in measurements at other incident beam energies.
The total $D^{\pm}$ cross section for 515~GeV/$c$ $\pi^-$-nucleon interactions
is $16.6\pm 4.5(stat) \pm 4.8(syst) \ \mu$b per nucleon. 

\acknowledgements
   
   We thank M.~Mangano and T.~Sj\"{o}strand for sharing 
their valuable insights on the 
phenomenology of charm hadroproduction and for helping us to
understand and utilize their programs. We also thank
S.~Kwan for several valuable discussions related to charm 
hadroproduction. We are grateful for the valuable contributions
of our colleagues on Fermilab experiment E672.
We thank the U.~S.\ Department of Energy, the National Science
Foundation, including its Office of International Programs, the
Universities Grants Commission of India, and Fermilab for their
support of this research.

\newpage

\begin{thebibliography}{10}
\bibitem{vogtscale}
The shapes of the calculated $p_T$ spectra are not entirely independent of the
  choices of the input parameters of the calculation. For illustrations of the
  sensitivity to the choice of PDF and/or the renormalization scale
  formulation, see Ref.\ \cite{vogt}.
\bibitem{e672-spec}
The diagram does not include the veto walls and steel absorber located upstream
  of the detector elements shown. The diagram also excludes the downstream muon
  identification system, which was located downstream of the detector elements
  shown in the diagram, and which was used primarily by Fermilab E672 to select
  events containing high mass dimuons.
\bibitem{muon-prot}
Scintillation counter veto walls surrounding a large steel absorber located at
  the front end of the Meson West spectrometer tagged muons that were produced
  upstream of the experimental hall.
\bibitem{dptrig}
Note that both the SINGLE LOCAL HI trigger and the LOCAL$\otimes$GLOBAL HI
  trigger are sensitive to high $p_T$ direct photons or photons from high $p_T$
  $\pi^0$ decays, but the LOCAL$\otimes$GLOBAL HI trigger exhibits greater
  sensitivity to high $p_T$ mesons that decay into more widely separated
  photons, such as $\eta\rightarrow\gamma\gamma$ or
  $\omega\rightarrow\pi^0\gamma$.
\bibitem{corr}
Corrections were applied to the measured $\Delta X$, $\Delta Y$, and
  $\Delta\theta_Y$ to account for the location of the center of curvature of
  the path of the charged particles, the change in $p_{\rm Z}$, and for the
  effects of the fringe field of the analysis magnet.
\bibitem{par-tune}
This particular {\sc pythia} parameter is stored in the array element PARP(81),
  and it was reduced from 1.3 GeV to 0.7 GeV. See the {\sc pythia} reference
  manual for additional details\cite{pylund}.
\bibitem{alpha}
The $\alpha$ determined from the $D^{\pm}$ mesons in this data sample was
  $1.28\pm 0.33$. See Ref.\ \cite{blusk} for details.
\bibitem{xserr}
The two contributions to the uncertainties in the numbers of $D^{\pm}$ mesons
  quoted in Tables \ref{signal-allxf-table} and \ref{signal-posxf-table} are
  treated as uncorrelated statistical uncertainties in the differential cross
  section analyses. The two contributions to the uncertainties associated with
  each of the efficiencies are treated as uncorrelated systematic uncertainties
  in the differential cross section analyses.
\bibitem{mangano_p}
The NLO pQCD calculations were performed using a set of programs provided by M.
  Mangano. (See Ref.\ \cite{mangano4} and references therein). Unless otherwise
  explicitly stated, the theoretical parameters were set to the defaults
  specified in Ref.\ \cite{mangano4}.
\bibitem{xffac}
The factor of 1.6 is from a NLO pQCD calculation, and therefore does not
  account for the leading particle effect which results in enhanced forward
  production of charm hadrons whose light quark is a spectator valence quark
  from the incoming beam particle.
\end{thebibliography}

\newpage
\noindent
\mediumtext
\begin{table}
\caption{Numbers of $D^{\pm}$ candidates and associated efficiencies per 
$p_T$ bin for $D^{\pm}$ mesons with $x_F>-0.2$.
For the number of $D^{\pm}$ candidates,  
the first uncertainty represents the statistical error in 
the total number of combinations in the $D^{\pm}$ signal region,
while the second uncertainty  
was determined by varying the fitted background shapes and the 
range over which the background fits were performed.  
The quoted uncertainties associated with the efficiencies
reflect the statistical and systematic errors, respectively.}
\label{signal-allxf-table}
\begin{tabular}{cccc}
\multicolumn{1}{c}{$p_T$ bin (GeV/$c$)} 
&\multicolumn{1}{c}{Number of $D^{\pm}$}
&\multicolumn{1}{c}{Trigger Eff. (\%)} 
&\multicolumn{1}{c}{Fitted Recon. Eff. (\%)} \\
\tableline
1--2&$43\pm 12\pm 6$&$0.0138\pm 0.0005\pm 0.0023$&$8.9\pm 0.6\pm 0.9$ \\
2--3&$43\pm 9\pm 4$&$0.0559\pm 0.0028\pm 0.0093$&$14.4\pm 0.9\pm 1.5$  \\
3--4&$16\pm 5\pm 2$&$0.229 \pm 0.018\pm 0.038$&$16.4\pm 1.2\pm 1.6$  \\
4--6&$6\pm 3\pm 2$&$1.49\pm  0.13\pm 0.25$&$17.4\pm 1.5\pm 1.7$ \\
6--8&$2\pm 1.4$&$8.81\pm 2.2 \pm 1.5$&$17.6\pm 3.7\pm 1.8$ \\
\end{tabular}
\end{table}

\mediumtext
\begin{table}
\caption{Numbers of $D^{\pm}$ candidates and associated efficiencies per 
$p_T$ bin for $D^{\pm}$ mesons with $x_F>0$. 
For the number of $D^{\pm}$ candidates,  
the first uncertainty represents the statistical error in 
the total number of combinations in the $D^{\pm}$ signal region,
while the second uncertainty  
was determined by varying the fitted background shapes and the 
range over which the background fits were performed.  
The quoted uncertainties for the efficiencies represent 
the statistical and systematic errors, respectively.}
\label{signal-posxf-table}
\begin{tabular}{cccc}
\multicolumn{1}{c}{$p_T$ bin (GeV/$c$)} 
&\multicolumn{1}{c}{Number of $D^{\pm}$}
&\multicolumn{1}{c}{Trigger Eff. (\%)}
&\multicolumn{1}{c}{Fitted Recon. Eff. (\%)} \\ 
\tableline
1--2&$41\pm 10\pm 5$&$0.0141\pm 0.0006\pm 0.0023$&$11.3\pm 0.8\pm 1.1$ \\
2--3&$31\pm 7\pm 4$&$0.0581\pm 0.0033\pm 0.0096$&$14.6\pm 1.1\pm 1.5$  \\
3--4&$13\pm 4\pm 2$&$0.222 \pm 0.020\pm 0.037$&$16.5\pm 1.5\pm 1.7$  \\
4--6&$5\pm 3\pm 2$&$1.36\pm  0.15\pm 0.23$&$17.8\pm 1.8\pm 1.8$ \\
6--8&0&$9.72\pm 2.8 \pm 1.6$&$18.5\pm 4.1\pm 1.9$ \\
\end{tabular}
\end{table}

\mediumtext
\begin{table}
\caption{Inclusive $D^{\pm}$ differential cross section per nucleon in 
515~GeV/$c$ $\pi^-$-nucleon interactions integrated over $D$ meson $x_F>-0.2$
and $x_F>0$.  The quoted uncertainties 
in the cross sections reflect the statistical and systematic errors, 
respectively. See the text for additional details.}
\label{xs-table}
\begin{tabular}{cccc}
$p_T$ bin &$p_T^{lw}$&
${d\sigma\over dp_T} (\mu {\rm b/(GeV}/c))$&
${d\sigma\over dp_T} (\mu {\rm b/(GeV}/c))$ \\
(GeV/$c$) & (GeV/$c$) &$x_F>-0.2$&$x_F>0$ \\
\tableline
1--2&1.44$\pm$0.02&$4.9\pm 1.5\pm 1.2$&$3.6\pm 1.0\pm 0.9$ \\
2--3&2.41$\pm$0.02&$0.75\pm 0.17\pm 0.19$&$0.52\pm 0.13\pm 0.13$ \\
3--4&3.41$\pm$0.02&$0.060\pm 0.020\pm 0.016$&$0.050\pm 0.017\pm 0.014$ \\
4--6&4.70$\pm$0.05&$0.0016\pm 0.0010\pm 0.0004$&$0.0015\pm 0.0010\pm 0.0004$ \\
6--8&6.75$\pm$0.05&$0.000091\pm 0.000064\pm 0.000037$&- \\
\end{tabular}
\end{table}

\newpage

\end{document}